\documentclass[prd,twocolumn,showpacs,preprintnumbers,amsmath,amssymb,nofootinbib]{revtex4}

\usepackage{graphicx}
\usepackage{dcolumn}
\usepackage{bm}

\catcode`\@=11
\def\lsim{\mathrel{\mathpalette\@versim<}}
\def\gsim{\mathrel{\mathpalette\@versim>}}
\def\@versim#1#2{\vcenter{\offinterlineskip
\ialign{$\m@th#1\hfil##\hfil$\crcr#2\crcr\sim\crcr } }}
\catcode`\@=12

\newcommand{\nn}{\nonumber}

\newcommand{\be}{\begin{equation}}
\newcommand{\ee}{\end{equation}}
\newcommand{\bea}{\begin{eqnarray}}
\newcommand{\eea}{\end{eqnarray}}
\newcommand{\ben}{\begin{enumerate}}
\newcommand{\een}{\end{enumerate}}
\newcommand{\bit}{\begin{itemize}}
\newcommand{\eit}{\end{itemize}}

\begin{document}
\input epsf.tex


\preprint{
\parbox{35mm}{
KUNS-2014 \\
KANAZAWA-06-02
}
}
\title{
Fine-tuning in gauge mediated supersymmetry
breaking models and induced top Yukawa coupling
}
\author{Tatsuo Kobayashi}
\email{kobayash@gauge.scphys.kyoto-u.ac.jp}
\affiliation{
Department of Physics, Kyoto University, 
Kyoto 606-8502, Japan
}%
\author{Haruhiko Terao}
\email{terao@hep.s.kanazawa-u.ac.jp}
\author{Akito Tsuchiya}
\email{tsucchi-@hep.s.kanazawa-u.ac.jp}
\affiliation{
Institute for Theoretical Physics, Kanazawa
University, Kanazawa 920-1192, Japan
}%
\date{\today}
\begin{abstract}
It is shown that fine-tuning of the Higgs parameters
stronger than a few \% is required at the best in the models 
with gauge mediated supersymmetry breaking.
With the aim of solving this problem, 
we consider a new type of models in which the top
Yukawa coupling is induced at TeV scale through mass
mixing with unknown matter fields.
Then it is found that the fine-tuning problem can be
eliminated essentially.
We discuss some phenomenological features of this model
and also consider the extension to the next-to-minimal 
models.

\end{abstract}



\maketitle

\section{Introduction}

The supersymmetric extension of the standard model (SM)
is the most promising candidate for new physics
beyond TeV scale. Among many attractive features, the 
gauge coupling unification in the supersymmetric models
is remarkable.
However the fact that the Higgs boson besides any 
superparticle has not been discovered by the LEP2 
experiment, casts the so-called ``little hierarchy problem''
\cite{BG,finetune} on this direction. 

The lower experimental 
bound of the lightest Higgs boson mass,
which is $114$GeV\cite{LEPII},
indicates that the mass should be
lifted up by a large radiative correction of the
top quark loop\cite{massbound,2loop}.
{}For the minimal supersymmetric standard model (MSSM), 
the mass of top squark (stop), $m_{\tilde{t}}$, is
required to be much larger than the weak scale in general.
The Higgs mass bound, which is also 
rather sensitive to the trilinear
coupling of top squarks and Higgs boson, $A_t$,
leads 
\be
\begin{array}{lll}
m_{\tilde{t}} \geq 500\mbox{GeV} & \mbox{for} &
|A_t| \sim |m_{\tilde{t}}|, \\
m_{\tilde{t}} \geq 1\mbox{TeV} & \mbox{for} &
|A_t| \ll |m_{\tilde{t}}|,
\end{array}
\label{stopmass}
\ee
for a large value of $\tan \beta$. 
Moreover this mass bound increases rapidly,
as $\tan \beta$ decreases.
This ``hierarchy'' between the weak scale and the 
supersymmetry breaking mass scale turns out to bring
about a severe fine-tuning problem as follows.

The quartic couplings of the Higgs fields are related
with the gauge couplings in the MSSM.
Therefore the scale of electro-weak symmetry 
breaking (EWSB) is determined solely by the mass 
parameters of the Higgs bosons.
{}For a moderate value of $\tan \beta$, such 
relation is simply given by
\be
\frac{M_Z^2}{2} = - \mu^2 - m_{H_u}^2,
\label{MSSM}
\ee
where $\mu$ denotes the supersymmetric mass of the Higgs
multiplets (or the higgsino mass) and $m_{H_u}^2$
denotes the soft supersymmetry breaking mass (soft scalar mass)
of the up-type Higgs boson $H_u$.
This relation indicates that both $\mu$ and $|m_{H_u}|$
should be of the order of the weak scale, otherwise
very accidental cancellation is required between them. 

However, the soft scalar mass $m_{H_u}^2$ receives
a fairly large radiative correction due to the
very heavy stop mass and the large top Yukawa
coupling $y_t$.
Supersymmetry protects scalar masses from quadratic
corrections, but not from logarithmic corrections.
The correction to  $m_{H_u}^2$ at the one-loop level
is given approximately by
\be
\delta m_{H_u}^2 \sim - \frac{3}{4\pi^2}y_t^2
m_{\tilde t}^2 \ln \frac{\Lambda_{MSSM}}{m_{\tilde t}},
\label{log-h-mass}
\ee
where $\Lambda_{MSSM}$ denotes the up-most scale
of the MSSM. 
If we take this scale to be the scale of the grand
unified theory (GUT), 
$\Lambda_{GUT} \sim 2 \times 10^{16}$GeV, then
this correction is estimated as 
$|\delta m_{H_u}^2| \geq (700\mbox{GeV})^2$ for
$m_{\tilde{t}} \geq 500$GeV.
Indeed this negative correction to the Higgs soft mass
induces the radiative EWSB \cite{radSB}, 
which is one of the beautiful features of the
supersymmetric models.
Now, however, it is found that the size of the 
correction is too large.
The realistic EWSB does not occur in the
MSSM without delicate fine-tuning,
since the Higgs mass parameters $\mu$ and $m_{H_u}$
are not related with each other in origin. 
Therefore, we may expect that the correction to the
Higgs soft mass is suppressed for some reason.

Somewhat small sizes of the mass parameters 
$\mu$ and $|m_{H_u}|$ are favorable from the
point of view of the dark matter as well.
In the supersymmetric theories, the
lightest superparticle (LSP), which is stable due to
the R-parity symmetry,
is supposed to be a good candidate for 
the dark matter.
However, if the higgsino mass $\mu$ is pretty larger than
the bino mass, then the dark matter relic density
far exceeds the observed amount.
On the other hand, when $\mu$ is comparable with the bino mass,
the LSP is composed of a bino-higgsino mixture, which can
lead to a suitable amount of the relic density \cite{darkmatter}.

Thus it would be worthwhile to ask what kinds of 
models instead of the MSSM offers us more natural
frameworks. 
The gauge mediated supersymmetry breaking (GMSB) models 
\cite{GMSB}
may be thought as the good candidate,
since the uppermost scale $\Lambda_{MSSM}$ can be lowered
to about $50$TeV.
Then the large logarithmic factor in (\ref{log-h-mass})
is reduced to 4$\sim$5.
However it has been known that the fine-tuning 
in the GMSB models is not improved, rather
becomes worse because of the following reasons.

In the GMSB framework, the messenger fields, which acquire
the soft scalar masses through interaction with the 
dynamical supersymmetry breaking (DSB) sector, decouples at
the messenger scale $\Lambda_{mess}$.
Then the ratio among the various soft supersymmetry
breaking parameters induced at $\Lambda_{mess}$
are given uniquely by the SM gauge couplings.
In the case of the low scale GMSB with 
{\it e.g.} $\Lambda_{mess} \sim 50$TeV, 
the ratio of soft masses of the stop $m_{\tilde{t}}$
and of the right-handed selectron $m_{\tilde{e}}$
may be evaluated roughly as
\be
\frac{m_{\tilde{t}}^2}{m_{\tilde{e}}^2}
\simeq \frac{(4/3)g_3^4}{(3/5)g_1^4}
\simeq (7 \sim 8)^2.
\ee
Meanwhile, the slepton mass is bounded experimentally
as $m_{\tilde{e}} \geq 100$GeV.
Therefore $m_{\tilde{t}}$ appears to be quite large,
$m_{\tilde{t}} \geq 700$GeV, in the low scale GM.
Such a large stop mass makes the correction to the
Higgs soft mass even bigger in spite of the smaller
logarithmic factor \cite{AG,dFM}.

It should be also noted that A-parameters are not induced
or very suppressed at the messenger scale, except for 
special cases with vector messengers at the GUT scale. 
Therefore the tri-linear coupling $A_t$ remains small 
up to low energy in the low scale GMSB.
Then the Higgs mass bound imposes a very severe constraint
on the stop mass $m_{\tilde{t}}$.

It has been also known that the GMSB has a difficulty to 
lead the proper size of $B$-parameter as well as the
$\mu$-parameter \cite{muproblem,GMSB}.
The easiest solution escaping this $\mu$-problem would be
to consider the next-to-MSSM (NMSSM) type extension \cite{NMSSM}.
There we introduce a SM gauge singlet $N$ and consider
the superpotential
\be
W = \lambda_H N H_u H_d + \cdots.
\label{NMSSM}
\ee
The $\mu$-parameter is given by the non-vanishing
vacuum expectation value (vev) of $N$ as 
$\mu = \lambda_H \langle N \rangle$.
Then the  B-parameter in the MSSM is just given by the
trilinear coupling $A_H$ of the scalar fields,
\be
V = - \lambda_H A_H N H_u H_d + \cdots.
\label{AH}
\ee
The problem is that $\tan \beta$ is linked to the soft scalar
mass of the singlet $N$ and cannot be chosen freely
contrary to  the MSSM.
In practice, it will be seen later
that $\tan \beta$ is restricted to be small in the
realistic models \cite{dFM}.
Then the experimental Higgs mass bound leads to a
very strong constraint on the soft scalar mass 
$m_{\tilde{t}}$ in general.

So far various approaches have been proposed aiming to
solve this ``little hierarchy problem''
\cite{
Casas,fathiggs,Delgado,supersoft,superLH,
KT,KNT,NPT,KKLT,DK,KN,twinhiggs}.
Many of them push up the Higgs mass bound by introducing
other new interactions and assume a fairly small
$m_{\tilde{t}}$ to suppress the correction to the Higgs
mass.
The GMSB seems inappropriate in these approaches,
since the mass bound for the selectron 
leads to a large stop mass anyway. 
Apart from the GMSB, it is also a way to consider special
setup generating
the soft breaking parameters at some high energy scale
such that $|m_{H_u}|$ appears to be small enough at 
the weak scale,
as {\it e.g.} the focus point solution \cite{focus},
though the ground leading to the special supersymmetry
breaking  parameters assumed there is missing.
Another interesting scenario is the mixed modulus-anomaly 
mediation, the so-called mirage mediation.
In a certain class of such models the logarithmic 
correction (\ref{log-h-mass}) 
is cancelled by the anomaly mediation effect, 
and that leads to the favorable spectrum that 
stop is heavy and the soft scalar mass $|m_{H_u}|$ is small, 
while $|A_t|$ is large \cite{KKLT}.
The recently proposed setup with a negative 
$m_{\tilde{t}}^2$ at the GUT scale \cite{DK}
would be also an
interesting possibility in this point of view.

Another type of approach is to protect the Higgs
mass from the large radiative corrections.
The well-known examples of this kind are
the (supersymmetric) little Higgs models
\cite{superLH}, where the
Higgs scalar is the pseudo NG boson and the logarithmic
divergence is suppressed by two loop factors.
Two of the authors have also proposed the models, in which
the Higgs mass correction is made even finite, {\it i.e.}
cutoff independent, by use of the superconformal dynamics
\cite{KT,KNT}.
In these approaches a large $m_{\tilde{t}}$ is also
admitted and, therefore, the GMSB can be incorporated
in the models.

In this paper we would like to consider this problem
in relation to the origin of  prominently large top
quark mass.
It is obvious that the correction to the Higgs mass 
is not so enhanced, if the top Yukawa coupling $y_t$
is small above TeV scale.
Here we consider the models in which the top Yukawa coupling
is induced effectively at TeV scale through mixing between
top quarks and extra vectorlike particles \cite{KNT}.
In order to get the top Yukawa coupling large enough,
the Higgs field is assumed to be coupled with the
extra particles  rather strongly.

Furthermore we consider the ``partial GMSB'' to protect
the Higgs soft mass from a large loop correction of the extra
particles. The gauge group of the model is assumed to be
a direct product of the SM gauge group $G_{SM}$ and
the extra gauge group $G_{ex}$. 
We suppose also that the extra vectorlike fields are 
neutral to the color $SU(3)$, while the gauge messengers 
are neutral to $G_{ex}$.
Then all other scalar masses than the squark masses;
the soft scalar masses of the Higgs, the sleptons and the 
extra particles, 
become small automatically.
The gluino is also heavy as in the ordinary GMSB.
The point is that the squark masses do not enhance the 
Higgs mass, since the top Yukawa coupling is assumed to 
be small above TeV scale.
It may be also said that the large logarithmic correction to 
the Higgs mass by the stop loop is cutoff effectively 
at TeV scale.
Consequently, our model realizes the fairly small
Higgs mass parameters at the weak scale
in spite of the large stop masses,
and, therefore, is released from the fine-tuning problem.

This paper is organized as follows.
In section II we examine the degree of fine-tuning
in the GMSB models with taking account of the Higgs
mass bound by LEP2. It will be seen that the delicate
fine-tuning within a few \% level is required at the very
best.
In section III we present an explicit model with the
induced top Yukawa coupling and discuss some 
phenomenological features.
Then we quantify the degree of fine-tuning required in the
new model explicitly in section IV.
In section V we consider the extension to the NMSSM type
in order to solve the $\mu$-problem. 
The last section is devoted to the conclusion.

\section{Fine-tuning in the GMSB models}

We consider the minimal GMSB model with vectorlike
chiral messenger fields $(q, l)$ and $(\bar q, \bar l)$. 
The fields $q$ and $l$
transform as
$(\bf{3}, \bf{1}, -1/3)$ and 
$(\bf{1}, \bf{2}, 1/2)$ under the SM gauge group
$G_{SM} = SU(3)_c \times SU(2)_W \times U(1)_Y$ 
respectively,
and $(\bar{q}, \bar{l})$ carry their conjugate 
representations.
These fields do not have any direct interactions with the 
MSSM sector.
Since the fields $q$ and $l$ are combined into
a $\bf{5}$ representation of the $SU(5)$ 
group, the gauge
coupling unification is maintained.

In the minimal GMSB model, the messengers couple with
the DSB sector through the superpotential given by
\be
W_{GMSB} = \kappa_q S q \bar{q} + \kappa_l S l \bar{l},
\ee
where $S$ is a SM gauge singlet.
The dynamics of the DSB sector is supposed to generate
non-vanishing vevs for $S$ and the F-component $F_S$
at the same time.
These messenger decouples from the MSSM sector at the
messenger scale 
$\Lambda_{mess} \sim \kappa \langle S \rangle$.

The F-component vev induces soft supersymmetry breaking
parameters in the MSSM sector as finite loop 
corrections by the messenger fields.
The formula for the induced parameters are given 
at the leading order by
\bea
M_a &=& \frac{\alpha_a}{4\pi} B_S~~~(a=1,2,3),  
\label{GMformula1} \\
m^2_{\tilde{f}_i} &=& 2 
\left[
C_{3i} \left( \frac{\alpha_3}{4 \pi} \right)^2
+ C_{2i} \left( \frac{\alpha_2}{4 \pi} \right)^2 
\right.
\nn \\
& &
\left.
+ \frac{3}{5}Y_i^2 \left( \frac{\alpha_1}{4 \pi} \right)^2
\right]
B_S^2,  
\label{GMformula2} \\
A_i &=& 0,
\label{GMformula3}
\eea
where $B_S = \langle F_S \rangle / \langle S \rangle$, and
$C_2, C_3, Y$ denote the quadratic Casimirs of the representation
for $SU(2)_W$ and $SU(3)_c$ and the hypercharge for
$U(1)_Y$ respectively.
It is a general property that the induced A-parameter is 
very small at the messenger scale in the chiral messenger
GMSB. Although some special models of GMSB by vector messengers
may have sizable A-parameters \cite{GR}, 
we do not consider such cases in this paper.

Here we note that all the soft scalar masses are fixed up to
the overall scale of the supersymmetry breaking parameter $B_S$,
which is also bounded as discussed below.
Therefore we can tune only the $\mu$ parameter in the GMSB
to satisfy the  minimization condition for the Higgs 
potential,
\be
\frac{M_Z^2}{2} = - \mu^2 + 
\frac{m^2_{H_d} - m^2_{H_u} \tan^2 \beta}{\tan^2 \beta -1}.
\label{MSSM2}
\ee
The degree of fine-tuning may be defined by
\be
\Delta_{\mu^2}(M_Z^2) = \left|
\frac{\mu^2}{M_Z^2}
\frac{\partial M_Z^2}{\partial \mu^2}
\right|
= 2 \frac{\mu^2}{M_Z^2}.
\label{degree}
\ee

\begin{widetext}
These soft mass parameters are subject to the renormalization
group (RG) of the MSSM below the messenger scale.
The soft mass of the right handed selectron $m_{\tilde{e}}$
at the weak scale is given roughly by
\be
m_{\tilde{e}}^2 \sim \frac{6}{5} 
\left( \frac{\alpha_1}{4 \pi} \right)^2 B_S^2
+ \frac{2}{11} \left[
\left( \frac{\alpha_1(\Lambda_{mess})}{\alpha_1(M_Z)} 
\right)^2 -1
\right] M_1^2(M_Z)
- M_Z^2 \sin^2 \theta_W \cos 2 \beta,
\label{selectron}
\ee
where $\theta_W$ denotes the Weinberg angle and 
$\sin^2 \theta_W \simeq 0.232$.
The experimental lower bound 
$m_{\tilde{e}} \geq 100$GeV
constrains the supersymmetry breaking parameter $B_S$
to be more than roughly $50$TeV. 
Therefore it would be reasonable to take the messenger
scale $\Lambda_{mess} > 50$TeV for the typical
low scale GMSB, provided the DSB sector induces
$\langle F_S \rangle \leq \langle S \rangle^2$
\footnote{
In the case of $\langle F_S \rangle \sim \langle S \rangle^2$,
the formulae (\ref{GMformula1}), (\ref{GMformula2})
are modified with a factor of $O(1)$ \cite{GMSB}.
}.

In such a low scale GMSB case, the logarithmic factor 
in the one loop correction given by (\ref{log-h-mass})
is somewhat reduced,  
$\log(\Lambda_{mess}/m_{\tilde{t}}) \sim 5$.
However, the soft scalar mass $m_{\tilde{t}}$ is quite
large instead. In practice it is found by solving the 
one loop RG equations that 
$m_{\tilde{t}}/m_{\tilde{e}} \sim 7$
at the weak scale.
Therefore $m_{\tilde{t}} > 700$GeV is required from the
mass bound to (\ref{selectron}) for large $\tan \beta$.
Then the soft mass of the Higgs turns out to be 
$m_{H_u}^2 < -(300\mbox{GeV})^2$ at the weak scale
by the one-loop RG analysis.
Then this result shows that about 5\% fine-tuning for
the $\mu^2$ parameter is required at the best \cite{AG,dFM}.
{}For a small value of $\tan \beta$, 
the required stop mass becomes
slightly bigger, therefore the fine-tuning is worse. 

In practice, however, the experimental Higgs mass bound,
$m_{h^0} > 114.4$GeV, is found to put a much stronger
constraint than the selectron mass bound does.
Within the two loop approximation, 
the analytic expression for the mass bound 
has been given by \cite{2loop}
\be
m_{h^0}^2 \leq
M_Z^2 \cos^2 2 \beta
\left(
1 - \frac{3 m^2_t}{8\pi^2 v^2}t
\right) 
+  \frac{3 m_t^4}{4\pi^2 v^2}
\left[
\frac{1}{2} \tilde{X}_t + t +
\frac{1}{16 \pi^2}
\left(
\frac{3m^2_t}{2v^2} - 32 \pi \alpha_3
\right)
\left(
\tilde{X}_t t + t^2
\right)
\right],
\ee
where $t = \log{m^2_{\tilde{t}}/m^2_t}$ defined 
with the average stop mass 
$m^2_{\tilde{t}} = 
\sqrt{
m^2_{\tilde{Q}_3} 
m^2_{\tilde{\bar{u}}_3}
}$ 
($m^2_{\tilde{Q}_3}$ and $m^2_{\tilde{\bar{u}}_3}$
denote the soft scalar masses for the left handed
stop-sbottom doublet and the right handed stop
respectively.),
$v=173.7$ GeV is the vev of Higgs and
$m_t$ is the on-shell running mass computed from
the physical top mass $M_t = 172.7$GeV,
i.e. $m_t= 164.5$GeV.
The factor $\tilde{X}_t$ stands for the mixing between
the left and right handed stops and is given explicitly
by
\be
\tilde{X}_t =
\frac{2 \tilde{A}^2_t}{m^2_{\tilde{t}}}
\left(
1 - \frac{\tilde{A}^2_t}{12 m^2_{\tilde{t}}}
\right),
\label{mixing}
\ee
where $\tilde{A}_t = A_t - \mu \cot \beta$.
In this analytic expression,
the maximal value of mixing is $\tilde{X}_t = 6$,
which is realized for
$\tilde{A}_t/m_{\tilde{t}} = \sqrt{6}$.
The constraint on $m_{\tilde{t}}$ imposed
by the Higgs mass bound is rather sensitive to the 
stop mixing and $m_{\tilde{t}}$ is required to be
more than 1TeV for a small $A_t$.
\end{widetext}

In the GMSB, in which $A_t = 0$ at the messenger scale 
$\Lambda_{mess}$,
the value of $A_t$ at the weak scale is 
given approximately by
\be
A_t \sim - \frac{8}{3\pi} \alpha_3 M_3 
\log \left(\frac{\Lambda_{mess}}{m_{\tilde{t}}}\right).
\label{At}
\ee
Therefore, $|A_t|$ is rather small for the low 
scale GMSB and  $|A_t| \leq m_{\tilde{t}}$ for any messenger
scale $\Lambda_{mess} < \Lambda_{GUT}$.
This is a crucial feature of the GMSB models for
the fine-tuning problem.

\begin{figure}[htb]
\includegraphics[width=0.45\textwidth]{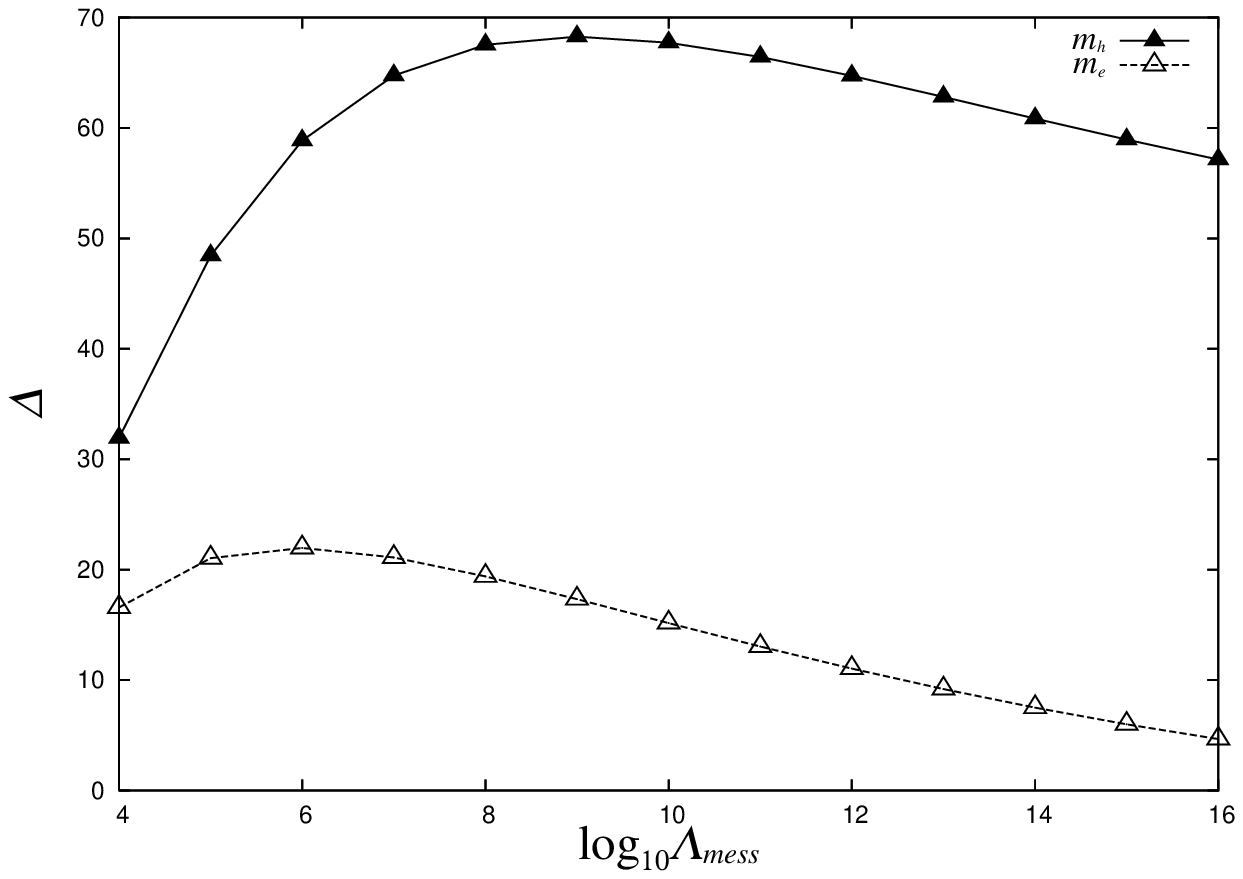}
\caption{\label{fig1} 
The minimum degrees of fine-tuning required in the MSSM
with $\tan \beta = 10$ are shown with respect to 
various messenger scales $\Lambda_{mess}$ of the GMSB.
The upper line is given by the Higgs 
mass bound, while the lower line is given by the
selectron mass bound.
}
\includegraphics[width=0.45\textwidth]{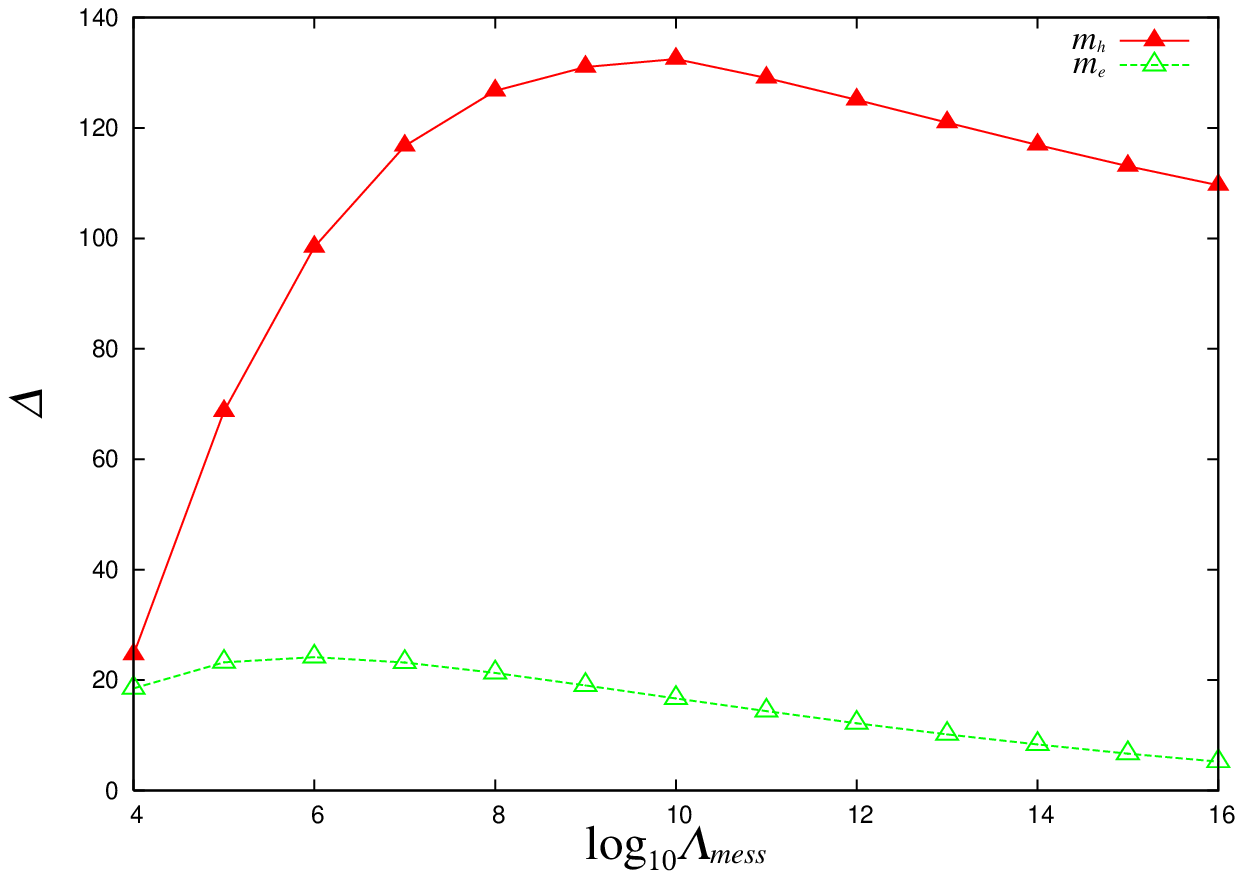}
\caption{\label{fig2} 
The minimum degrees of fine-tuning required in the MSSM
with $\tan \beta = 5$ are shown with respect to 
various messenger scales $\Lambda_{mess}$ of the GMSB.
The conditions for the lines are the same as in Fig.~1.
}
\end{figure}

In Fig.~1 and Fig.~2,  the minimum degrees of 
fine-tuning required in the MSSM  are shown 
in the cases with $\tan \beta = 10$ and $5$ respectively.
The messenger scale $\Lambda_{mess}$ is varied up to
the GUT scale. It is seen that the Higgs mass bound puts
a far stronger constraint on the fine-tuning than the
selectron mass bound.
Both the constraints become milder as the messenger
scale goes higher. 
However the fine-tuning stronger than 2\% is found to be
required at the very best.
For the case with $\tan \beta < 5$, the fine-tuning 
is severer than 1\% level except for the case with
a very low scale messenger scale.
The dependence of $\Delta_{\mu^2}$ on the $\tan \beta$
parameter in the case of a high messenger scale 
is shown in Fig.~3.
It is seen that the bound by the lightest
Higgs boson mass becomes extremely severe for a small 
$\tan \beta < 5$.
\footnote{
It is noted that the analyses of the degree 
of fine-tuning required
in the GMSB shown in Refs.~\cite{AG,dFM}
put emphasis on the slepton mass bound.
}

\begin{figure}[htb]
\includegraphics[width=0.45\textwidth]{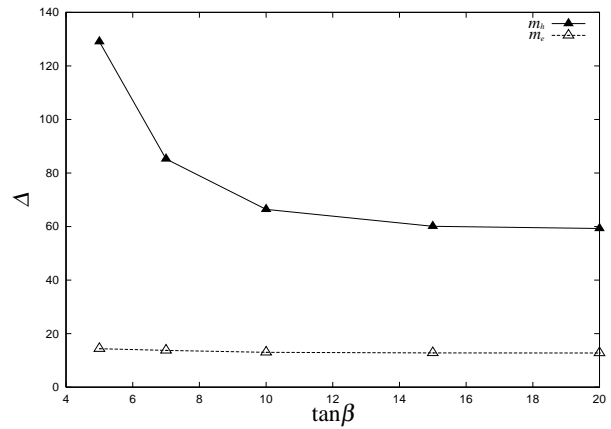}
\caption{\label{fig3} 
The minimum degrees of fine-tuning required in the MSSM
with the messenger scale $\Lambda_{mess}=10^{16}$GeV
are shown for various values of  $\tan \beta$. 
The conditions for the lines are the same as 
in Fig.~1.
}

\end{figure}

\section{Extra $SU(3)$ model with the induced top Yukawa 
coupling}

\subsection{The extra $SU(3)$ model}

In this section, we consider an explicit model 
avoiding this severe fine-tuning problem
within the GMSB framework.
The new model is based on the ideas of the ``induced top 
Yukawa coupling'' and the ``partial gauge mediation''.
Although the induced soft scalar mass of the Higgs 
field $H_u$ by the gauge messengers is not large,
the large soft mass transmitted through the stop
loop causes the problematic hierarchy.
If the top Yukawa coupling is as small as other
Yukawa couplings above TeV scale,
then this transmission can be effectively 
switched off.
Naturally all other scalar fields than squarks in the MSSM,
including the Higgs scalars, carry relatively small soft
scalar masses.

Meanwhile it is an interesting puzzle of the flavor physics
why only top quark is so heavy.
Here we take the picture that the large top Yukawa
coupling is induced at TeV scale dynamically.
There must be the additional sector coupled with
the Higgs field so strongly as to generate
the large top Yukawa coupling.
However, if the supersymmetry breaking is not mediated to the
new sector, then the Higgs field does not receive sizable
corrections in spite of the large coupling to the new sector. 
This is the reason why we apply the gauge
mediation acting partially. 

First we assume that the SM gauge group is extended
to $G= SU(3)_{ex} \otimes SU(3)'_C \otimes SU(2)_W
\otimes U(1)_Y$ and that
$SU(3)_{ex} \otimes SU(3)'_C$
is spontaneously broken to the color $SU(3)_C$
diagonally at a few TeV scale.
We introduce also vectorlike matter fields charged under
$G$ as
\be
\begin{array}{|c|ccccc|}
\hline 
 & SU(3)_{ex} & SU(3)'_C & SU(2)_W & U(1)_Y & ~~R~~ \\ \hline
~~\Phi~~ & {\bf 3} & {\bf 1} &  {\bf 2} &  1/6 & -  \\
\bar{\Phi} & {\bf 3}^* & {\bf 1} &  {\bf 2} &  -1/6  & - \\
\phi & {\bf 3} & {\bf 1} &  {\bf 1} &  2/3 & -  \\
\bar{\phi} & {\bf 3}^* & {\bf 1} &  {\bf 1} &  -2/3 & -  \\
\Omega & {\bf 3} & {\bf 3}^* &  {\bf 1} &  0 & +  \\
\bar{\Omega} & {\bf 3}^* & {\bf 3} &  {\bf 1} &  0 & + 
\\
\hline
\end{array}
\ee
where $R$ denotes the R-parity.
The superpotential of the model is given by 
\bea
& & W = W_{GMSB} + W_{MSSM} +  W_{yukawa} + W_{mass},  \\
& & W_{MSSM} = y_t Q_3 \bar{u}_3 H_u + \mu H_u H_d + \cdots, 
\label{WofMSSM}\\
& & W_{yukawa} = 
y_{\Phi} \Phi \bar{\phi} H_u + x Q_3 \Omega \bar{\Phi}
+ \bar{x} \phi \bar{\Omega} \bar{u}_3, \\
& & W_{mass} = M_{\Phi} \Phi \bar{\Phi}
+ M_{\phi} \phi \bar{\phi}
+ M_{\Omega} \Omega \bar{\Omega}. 
\label{Wofextra}
\eea
It is noted that the quarks not only of the third 
generation but also of an arbitrary combination of the
generations may couple to 
the extra fields in $W_{yukawa}$. 
Then such a combination of quarks turns out
to be heavy and is regarded as the top quark in the end.
We do not take account of the new couplings with $H_d$
either.  If the induced bottom Yukawa coupling is small
for some reason, then it will be also viable.

In the above superpotential,
the explicit mass terms for the extra fields 
in $W_{mass}$ are introduced by hand. 
To be explicit we simply set 
$M_{\Phi}=M_{\phi}=M_{\Omega}= 5$TeV hereafter,
though somewhat smaller mass parameters are also allowed 
by the EW precision experiments as is seen later on.
Then below $5$TeV, the extra sector decouples
from the MSSM sector.  
It is, indeed, a problem that these mass operators
can couple with the singlet $S$ with
non-vanishing F-component as well.
Actually this is the same problem
as the $\mu$-problem in the GMSB \cite{muproblem,GMSB}
and these couplings should be forbidden {\it e.g.}
by the discrete $U(1)_{PQ}$ symmetry.
We shall take care of these problems altogether in the
extension to the next-to-MSSM type discussed in 
section V. 
There it will be also shown that
the hierarchy between $\mu$ and the decoupling mass scale
can be generated naturally by the strong
gauge interaction of $SU(3)_{ex}$ \cite{KNT}.

We assume the top Yukawa coupling $y_t$ in the superpotential
given by (\ref{WofMSSM}) is as small as, {\it e.g.}, the
bottom Yukawa coupling. The explicit size of $y_t$ is irrelevant
in our discussions.
Rather, the large top Yukawa coupling 
is induced from $y_{\Phi}$ through
the mass mixing between the top quarks and the extra 
fields at the spontaneous symmetry breaking (SSB).
Hence $y_{\Phi}$ should be larger than 1 at the decoupling
scale. 
Indeed the extra sector is the same as the $SU(3)$ QCD
with 6 flavors, which belong to the so-called
conformal window \cite{seiberg}. 
Therefore the Yukawa coupling
as well as the extra gauge coupling satisfy an infrared
(IR) attractive fixed point \cite{KT,KNT}. 
In this case, the fixed point lies in the
strongly coupled region.
Thus such a large Yukawa coupling is realized naturally.
In the next section we shall discuss this in more detail.

The SSB of $SU(3)_{ex} \otimes SU(3)'_C$ to the color 
$SU(3)_C$ is brought about by 
non-vanishing vevs of $\Omega$ and $\bar{\Omega}$.
If the extra gauge coupling grows up around the decoupling
scale, then some strong dynamics may generate these vevs of
the TeV order.
Here we do not study on explicit dynamics leading
to the vevs, but rather assume 
\be
\langle \Omega^A_a \rangle =  \omega \delta^A_a,~~~~
\langle \bar{\Omega}_A^a \rangle =  \bar{\omega}  \delta_A^a,
\ee
where  $A$ and $a$ ($= 1, 2, 3$) 
denote indices of the 
fundamental representations of $SU(3)_{ex}$
and $SU(3)'_C$ respectively.
The vevs $\omega$ and $\bar{\omega}$ are also given
to be several TeV. 

After this SSB, the mass terms in the superpotential
(\ref{Wofextra}) are 
modified to 
\be
 M_{\Phi} \left(
\Phi + \frac{x \omega}{M_{\Phi}} Q_3
\right)\bar{\Phi}
+ M_{\phi}\phi 
\left(
\bar{\phi} + \frac{\bar{x} \bar{\omega}}{M_{\phi}} \bar{u}_3
\right).
\ee
These mass terms induce mixing between $SU(2)_W$ doublets,
$Q_3$ and $\Phi'$, and also singlets, $\bar{u}_3$ and
$\bar{\phi}$.
The mass eigenmodes $(Q_3', \Phi')$ and 
$(\bar{u}_3', \bar{\phi}')$ are given as 
\bea
\left( 
\begin{array}{c}
Q_3' \\ \Phi'
\end{array}
\right)
&=&
\left(
\begin{array}{cc}
\cos \theta_L & -\sin \theta_L \\
\sin \theta_L & \cos \theta_L 
\end{array}
\right)
\left(
\begin{array}{c}
Q_3 \\
\Phi
\end{array}
\right),
\\
\left(
\begin{array}{c}
\bar{u}_3' \\
\bar{\phi}'
\end{array}
\right)
&=&
\left(
\begin{array}{cc}
\cos \theta_R & -\sin \theta_R \\
\sin \theta_R & \cos \theta_R 
\end{array}
\right)
\left(
\begin{array}{c}
\bar{u}_3 \\
\bar{\phi}
\end{array}
\right),
\eea
where $\tan \theta_L = x \omega/M_{\Phi}$ and
$\tan \theta_R = \bar{x} \bar{\omega}/M_{\phi}$. 
These mixing angles are supposed to be of order one,
since the vevs $\omega$ and $\bar{\omega}$ are
assumed to be of the same order of the decoupling 
mass scale. 
The massless modes $Q_3'$ and $\bar{u}_3'$ are regarded
to be the third generation quarks in the effective MSSM.
The massive modes are
given by $(\Phi', \bar{\Phi})$ and of $(\phi, \bar{\phi}')$,
whose masses are slightly modified to
$M'_{\Phi}=M_{\Phi}/\cos \theta_L$ and
$M'_{\phi}=M_{\phi}/\cos \theta_R$.

After decoupling of the heavy particles, the theory is
reduced to be the MSSM with the additional
effective top Yukawa interaction,
\bea
W & \sim & W_{MSSM} \nn \\
& &
+ y_{\Phi} 
(\cos \theta_L \Phi'
-\sin \theta_L Q'_{3} )
(\cos \theta_R \bar{\phi}'
-\sin \theta_R \bar{u}'_{3} ) H_u \nn \\
& \sim & W_{MSSM} + y^{\rm eff}_t Q_3'\bar{u}_3'H_u.
\label{effsuperpot}
\eea
Here the induced top Yukawa coupling $y_t^{\rm eff}$
is  given explicitly by
\be
y^{\rm eff}_t  = y_{\Phi} \sin \theta_L \sin \theta_R.
\label{efftop}
\ee
This coupling can be so large as to lead the physical
top quark mass $M_t = 172.2$GeV, since $y_{\Phi}$
is rather large.

Next we consider the soft supersymmetry breaking 
parameters obtained in the effective theory.
We apply the same supersymmetry breaking sector and
the gauge messengers $(q, \bar{q})$ charged to the
$SU(3)'_C$, in place of the $SU(3)_C$,
and $(l, \bar{l})$.
Note that these messengers are neutral to the $SU(3)_{ex}$.
Therefore the gauginos of the $SU(3)_{ex}$ sector remain 
just massless all the way down to the low energy scale.
After the diagonal SSB, the gauge coupling of the color 
$SU(3)_C$ is given by 
$1/g^2_3 = 1/g^2_{ex} + 1/{g'_3}^2$.
Since the coupling $g_{ex}$ is fairly large, 
we have $g_3 \sim g'_3$.
Meanwhile the gaugino masses, $M_3$ for the 
$SU(3)_C$ and $M'_3$ for the $SU(3)'_C$,
at the SSB are related as
\be
\frac{M_3}{g^2_3} = \frac{M_{ex}}{g^2_{ex}}
+ \frac{M_3'}{{g'_3}^2},
\label{gaugino}
\ee
which leads to  $M_3 \sim M'_3$.
Therefore the gaugino mass spectra obtained at 
low energy is almost the same as the
conventional GMSB predicts.  

The induced soft scalar masses $m^2_{\Phi}$ and 
$m^2_{\phi}$ for the extra fields $\Phi$ and $\phi$, 
are small at the messenger scale in general, 
because these fields are neutral
to the $SU(3)'_C$. 
Besides, these soft masses are 
free from enhancement by the gaugino corrections.
As a result of this nature, the soft mass of the
Higgs scalar $m^2_{H_u}$ is not enhanced either
in spite of the large Yukawa coupling $y_{\Phi}$.
If we could switch off the gauge couplings of $SU(2)_W$
and $U(1)_Y$, then $m^2_{H_u}$ as well as
$m^2_{\Phi}$ and $m^2_{\phi}$ were all vanishing in this
partial GMSB. 
Thus the Higgs scalar can be ``sequestered''
from the large supersymmetry breaking sources.
In practice, the effects of the $SU(2)_W$
and $U(1)_Y$ gauge interactions are not negligible.
We postpone the quantitative analysis of the Higgs
mass and the degree of fine-tuning to section IV.

\subsection{Electro-Weak precision test}

In the extra $SU(3)$ model, the top quarks are mixed with
the heavy fermions with TeV scale masses. 
In general, such mixing effect induces sizable
corrections to the EW theory, while the corrections are
restricted rather strictly by the precision measurements
at the LEP experiments. 
Here we evaluate such corrections  expected in the
extra $SU(3)$ model explicitly.

First we evaluate the corrections to the EW
oblique parameters, $T$ and $S$. 
In the extra SU(3) model presented above, only the
up-type quarks couple with the extra matter fields.
Thus the custodial $SU(2)$ symmetry is largely broken
and, therefore, the correction to $T$-parameter, 
$\Delta T$, can be sizable.
Of course the corrections are more suppressed, as
the masses of the heavy fields are made larger by
their decoupling effect.
However we may find the lower bound for the decoupling
mass scale allowed by the EW precision test.

\begin{widetext}
Since the loop effects by the  scalar particles
are small due to the large soft masses,
we take only the fermionic fields into account.
To be explicit, we write the fermionic components
of the superfields as
\be
\begin{array}{ll}
Q'_{3L} \sim (t_L, b_L), ~~~~~ & 
\bar{u}'_{3R} \sim t_R, \\
\Phi'_L \sim (\chi_{L1}, \chi_{L2}), ~~~~~ & 
\phi'_R \sim \chi_{R}, \\
\Phi_R \sim (\chi_{R1}, \chi_{R2}), ~~~~~ & 
\phi_L \sim \chi_{L}.
\end{array}
\ee
These fermions also obtain the additional masses 
through the Yukawa interactions given in  
(\ref{effsuperpot}).
The Higgs scalar's vev,
$\langle H_u \rangle = (v_u, 0)$, for the EWSB
leads to the fermion mass terms given by
\be
\left(\bar{t}_L, \bar{\chi}_{L1}, \bar{\chi}_L \right)
\left(
\begin{array}{ccc}
y_{\Phi} s_L s_R v_u & 0 & 
- y_{\Phi} s_L c_R v_u \\
- y_{\Phi} c_L s_R v_u & M'_{\Phi} &
y_{\Phi} c_L c_R v_u \\
0 & y_{\Phi} v_u & M'_{\phi}
\end{array}
\right)
\left(
\begin{array}{c}
t_R \\
\chi_{R1} \\
\chi_R
\end{array}
\right) 
+ M'_{\Phi} \bar{\chi}_{L2} \chi_{R2},
\label{massmatrix}
\ee
where $s_{L(R)} = \sin \theta_{L(R)}$ and
$c_{L(R)} = \cos \theta_{L(R)}$.
This mass matrix shows that the EWSB induces
mixing between the weak doublets and singlets
of the left-handed fermions, $(t_L, \chi_L)$,
and also of the right-handed
fermions, $(t_R, \chi_{R1})$.
\end{widetext}

The additional oblique corrections are generated through
the weak doublet-singlet mixing given above. 
By calculating the one-loop diagrams with 
$M'_{\Phi}=M'_{\phi}=M'$, we may
evaluate $\Delta T$ as
\be
\Delta T \sim \frac{3}{16 \pi^2 \alpha_W}
\log \frac{{M'}^2}{m_t^2} 
\left(
\frac{y_{\Phi}v_u}{M'}
\right)^2
\sim O(1) \times \left(
\frac{y_{\Phi}v_u}{M'}
\right)^2,
\label{Tparameter}
\ee
unless the Yukawa coupling $y_{\Phi}$ is extremely large.
On the other hand, the experimental bound is  given roughly 
by $\Delta T \leq 0.1$ \cite{EWPT}.
Consequently, we find that a fairly low decoupling mass
$M'$ up to about $1$TeV is not excluded.
It is noted that the above oblique correction has the common 
or very similar structure with the top quark see-saw models
\cite{topseesaw} and some of the little Higgs models
\cite{LH}.

Incidentally, other constraints imposed by the precision
experiments are not restrictive.
The correction to $S$-parameter turns out to be relatively
smaller than $\Delta T$ by the one-loop calculation.
Another experimental bound discussed frequently is the
excess of the Z-boson decay, 
$R_b = \Gamma(Z \rightarrow b\bar{b})/
\Gamma(Z \rightarrow \mbox{hadrons})$.
However, as long as there is no (or a very small)
mixing mass for the bottom quarks in the
present model, the excess is also found to be
much less than the experimental bound ($\delta R_b < 1$\%).
These features are also common with the 
top quark see-saw models
\cite{topseesaw}.

\subsection{Sparticle spectra}

The extra fields do not have direct couplings with the
MSSM fields except for the Higgs field $H_u$.
The mass spectrum of these superparticles is 
determined in the GMSB, once the messenger scale 
$\Lambda_{mess}$ is fixed.
Meanwhile, the masses of the Higgs and the Higgsino 
are very dependent on the running behavior of the
Yukawa coupling $y_{\Phi}$,
which will be discussed in the next section.

However, the presence of the extra fields alter the
running behavior of the gauge couplings $g'_3, g_2, g_1$
of the $SU(3)'_C \otimes SU(2)_W \otimes U(1)_Y$.
The coefficients of the one-loop beta functions 
$\beta_a = (b_a/16\pi^2)g^3_a$ are given by
$(b_1,b_2,b_3) = (51/5,1,0)$.
Note that the gauge coupling of $SU(3)'_C$
is not asymptotically free but remains large for
any energy scale.
The gauge couplings for $SU(2)_W$ and $U(1)_Y$ both
increase at higher scale and cross $g'_3$ around
the GUT scale.
\footnote{
Unfortunately the gauge coupling unification is not
hold exactly, because the hypercharge assignment is
not fit for the $SU(5)$ representations.
Though it is interesting to construct viable models 
which can be embedded into a GUT theory, it is beyond
our present scope.
}

This change in the running gauge couplings influences
the supersymmetry breaking parameters obtained at the
low energy scale significantly. 
Fig.~4 shows the change of the supersymmetry breaking
mass spectra obtained at low energy of
the gaugino masses $M_a (a=1,2,3)$, the squark
masses $m_{\tilde{Q}_3}, m_{\tilde{\bar{u}}_3}$ 
and the right-handed
selectron mass $m_{\tilde{e}}$
with respect to the messenger scale $\Lambda_{mess}$.
As was mentioned just above, the gaugino mass $M_3$ is the 
same as $M'_3$ through the SSB. 
It is seen that the selectron turns out to be much 
heavier than the bino in the high energy gauge 
mediation, while both are light in the low energy 
mediation.
It will be found in the next section
that the masses of the  
stop particles should be as large as $1$TeV in 
order to satisfy the lightest Higgs mass bound.
Then the bino mass is given above the experimental
bound $M_1 >46$GeV.

\begin{figure}[htb]
\includegraphics[width=0.45\textwidth]{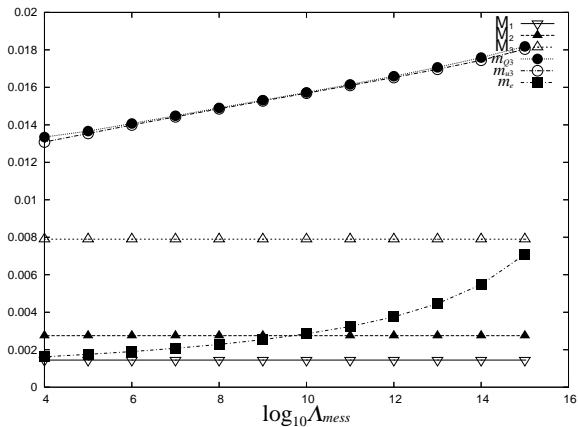}
\caption{\label{fig4} 
The supersymmetry breaking masses
($M_a (a=1,2,3)$: the gaugino masses,
$m_{\tilde{Q}_3}, m_{\tilde{\bar{u}}_3}$: 
the third generation squark masses,
$m_{\tilde{e}}$: the right-handed
selectron mass)
obtained at low energy for various
messenger scales $\Lambda_{mess}$
are shown. 
Here we take $B_S =\langle F_S \rangle/\langle S 
\rangle = 1$ as the unit.
}
\end{figure}

\section{Fine-tuning in the extra $SU(3)$ model}

\subsection{IR fixed point and running of $y_{\Phi}$}

In this section, we evaluate the degree of fine-tuning
required in the extra $SU(3)$ model explicitly, 
and see whether it 
is really improved as we expected in the previous section.
The fine-tuning is determined by the size of 
$|m^2_{H_u}|$ at the weak scale.
The induced value of $m^2_{H_u}$ by the gauge mediation
is rather changeable with respect to the messenger
scales, since the gauge couplings for $SU(2)_W$ and 
$U(1)_Y$ increase for the higher scale.
However, a more important point is the renormalization of 
$m^2_{H_u}$ below the messenger scale.
In practice, the renormalization is found to rely 
on the running behavior of the Yukawa coupling $y_{\Phi}$, 
although the scalar mass is not enhanced owing to the partial
GMSB.

We mentioned that there is an IR attractive fixed
point in the extra $SU(3)_{ex}$ gauge sector in
the previous section.
The couplings at the fixed point may be given as
$\alpha_{ex} = \alpha^*_{ex} = 27\pi/64$ and
$\alpha_{y_{\Phi}} = \alpha^*_{y_{\Phi}} =3\pi/8$
($\alpha_{ex} = g_{ex}^2/4\pi$,
$\alpha_{y_{\Phi}} = |y_{\Phi}|^2/4\pi$),
if we use the two-loop RG equations \cite{KNT}.
We find that the values of the fixed point couplings
are fairly large, though the perturbative 
approximation is not reliable there.
Even if these couplings are asymptotically free,
they grow up towards IR and approaches the
fixed point values.

In the followings, we consider two typical cases;
\begin{itemize}
\item[(1)] Both the couplings stay on the fixed point, 
$\alpha_{ex} = \alpha^*_{ex}$ and
$\alpha_{y_{\Phi}} = \alpha^*_{y_{\Phi}}$, at
all the scale below $\Lambda_{mess}$.
\item[(2)] Both the couplings $\alpha_{ex}$ and 
$\alpha_{y_{\Phi}}$ are given very small at
$\Lambda_{mess}$ and rapidly increase near 
TeV scale. 
\end{itemize}
We take the decoupling mass scale to be $5$TeV
in the analysis.
The supersymmetry breaking parameters at the weak
scale are evaluated by using the RG equations of the
MSSM with the induced top Yukawa coupling 
$y_t^{\rm eff} = 1.0$ below $5$TeV.

\subsection{Case (1)}

The gauge and Yukawa couplings in the extra sector
are set on their fixed point values.
It is noted that not only the gaugino mass $M_{ex}$
but also the A-parameter corresponding to the Yukawa
coupling $y_{\Phi}$, $A_{\Phi}$, are vanishing
\footnote{Though the A-parameter is not generated from 
the beginning  in the GMSB, it has been known that
the A-parameters decrease rapidly at lower scale
in the general superconformal field theories.}.
If we neglect the gaugino masses $M_1$ and $M_2$,
which give minute corrections indeed,
the one-loop RG equations for the scalar masses 
$m^2_{\Phi}$, $m^2_{\phi}$ and $m^2_{H_u}$ are given by
\be
\frac{d}{d\ln \Lambda}
\left( 
\begin{array}{c}
m^2_{\Phi} \\ 
m^2_{\phi} \\
m^2_{H_u} 
\end{array}
\right)
=
\frac{|y^2_{\Phi}|}{8\pi^2}
\left(
\begin{array}{ccc}
1~ & 1~ & 1 \\
2~ & 2~ & 2 \\
3~ & 3~ & 3 
\end{array}
\right)
\left( 
\begin{array}{c}
m^2_{\Phi} \\ 
m^2_{\phi} \\
m^2_{H_u} 
\end{array}
\right).
\label{focuspoint}
\ee
It is found that a general solution of this equation
is given by a superposition of two independent
constant modes and one mode decreasing very rapidly
towards IR.
Therefore the soft masses obtained at low energy
are always of the same order of their initial 
values, which are also independent of the 
messenger scale.

Here we should mention validity of this one-loop
analysis, since the fixed point couplings are not
very small.
It has been also shown in the all order of 
perturbation
\cite{sumrule}
that each beta function of the soft scalar mass is vanishing
at low energy and the masses satisfy the sum rule,
\be
m^2_{\Phi} + m^2_{\phi} + m^2_{H_u} \rightarrow 0.
\label{sumrule}
\ee
It is noted that the solutions of the above RG equations
also respect these properties.
Thus, it is certain that the soft masses at low energy
are independent of the messenger scale.
Indeed, the one-loop approximation is not totally reliable
quantitatively.
However, aspect of the soft masses is expected to be
seen by the one-loop RG equations.

Fig.~5 shows the actual running behavior of the soft scalar
masses, 
$m^2_{\Phi}/|m_{\Phi}|$, $m^2_{\phi}/|m_{\phi}|$ 
and $m^2_{H_u}/|m_{H_u}|$,
obtained by solving the one-loop RG equations for the 
extra $SU(3)$ model with a high messenger scale. 
Their initial values are given according to the GMSB
of this model and , therefore are fairly large.
The scale of the supersymmetry breaking parameters
are fixed so as to lead the lightest Higgs mass 
of the experimental bound, $114.4$GeV. 

\begin{figure}[htb]
\includegraphics[width=0.45\textwidth]{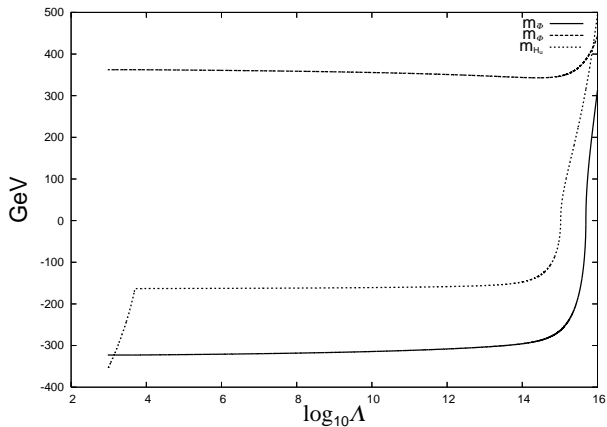}
\caption{\label{fig5} 
The RG flows of the soft scalar
masses, 
$m^2_{\Phi}/|m_{\Phi}|$, $m^2_{\phi}/|m_{\phi}|$ 
and $m^2_{H_u}/|m_{H_u}|$,
obtained by solving the one-loop RG equations for the 
extra $SU(3)$ model are shown in the case of
 a high messenger scale.
}
\end{figure}

It is seen that $m^2_{H_u}$ becomes negative quickly
just below the messenger scale and turns to be a constant.
The size of the Higgs scalar mass obtained at the
weak scale is not so small, since it receives another
sizable negative contribution below the decoupling
scale. 
For the models with a lower messenger scale,
the soft mass $|m^2_{H_u}|$ stabilizes at a much
smaller value.
Nevertheless, the lightest Higgs mass bound is more restrictive
due to the small A-parameter and makes
the fine-tuning worse.
As a result the fine-tuning of about 3\% is 
found to be still required even at the best. 
Thus we may say that the models of
Case (1) do not help to ameliorate the 
fine-tuning problem as expected naively.

\subsection{Case (2)}

Next we consider the case that both the couplings 
$\alpha_{ex}$ and $\alpha_{y_{\Phi}}$ increase near 
the TeV scale. 
To be explicit, we examine the case with 
$\alpha_{ex} \sim \alpha^*_{ex}$ and 
$\alpha_{y_{\Phi}} = 1/\pi$ 
$(y_{\Phi}=2)$ at the
decoupling scale as an example.
The RG flows of these couplings $g_{ex}$
and $y_{\Phi}$ are shown in Fig.~6.
In general, it is not necessary for the extra gauge 
coupling $\alpha_{ex}$ also increases near the decoupling
scale. The gauge coupling can approach the fixed point 
value at some higher energy scale. The required feature
is that the coupling $y_{\Phi}$ incerases near the 
decoupling scale. This may be realized with a small 
initial coupling properly chosen at the GUT scale
without recourse to fine-tuning.

Since the Yukawa coupling $y_{\Phi}$ becomes sizable
only at the fairly low energy scale, $10^{3-4}$GeV,
the Higgs field is almost completely decoupled
from the extra sector.
Therefore these soft scalar masses are driven only by
the gaugino contributions of the $SU(2)_W$ and
the $U(1)_Y$, which 
are rather small even at high energy scale.
The RG flows of the soft scalar masses
$m_{\Phi}$, $m_{\phi}, m^2_{H_u}/|m_{H_u}|$
are shown in Fig.~7.
The scale of the parameters are fixed so as to satisfy
the lightest Higgs mass bound again. 
It is seen that the soft mass of $m^2_{H_u}$ 
is almost constant
above the decoupling scale and receive the 
negative corrections through the Yukawa interactions
of $y_{\Phi}$.
Consequently, the low energy Higgs mass is found to be
small and can be comparable with the weak scale. 

As seen before, the lightest Higgs mass bound
is rather sensitive to the parameter $A_t$.
Note that the A-parameter $A_{\Phi}$ turns
out to be $A_t$ effectively after the SSB.
In practice, $A_{x}$, $A_{\bar{x}}$ and 
B-parameters for $\Phi$ and $\phi$ fields 
contribute to $A_t$ as well. Therefore we may
obtain a large $A_t$, if the B-parameters are
given large. This may be a good point of this senario,
since large A-parameters are not induced in the GMSM.
However we give an explicit evaluation of the 
fine-tuning only with $A_{\Phi}$ contribution,
since the values of these extra parameters are 
uncertain anyway.
The parameter $A_{\Phi}$ is generated below the
messenger scale with following the RG equation,
\be
\frac{d A_{\Phi}}{d \ln \Lambda} = 
\frac{3 |y_{\Phi}|^2}{4 \pi^2} A_{\Phi}
+ \frac{3 \alpha_2}{2 \pi} M_2
+ \frac{13 \alpha_1}{30 \pi} M_1. 
\label{Aphi}
\ee
In Fig.~7, the RG flow of $A_{\Phi}$ is also shown.
Below 5 TeV the flow line is connected with the
flow of $A_{t}$ in the MSSM.
It is seen that $A_{\Phi}$, or $A_t$, 
obtained at the weak scale is rather small 
even for the high energy GMSB.
Nevertheless the lightest Higgs mass bound
can be satisfied, since the stop particles are
so heavy. 

\begin{figure}[htb]
\includegraphics[width=0.45\textwidth]{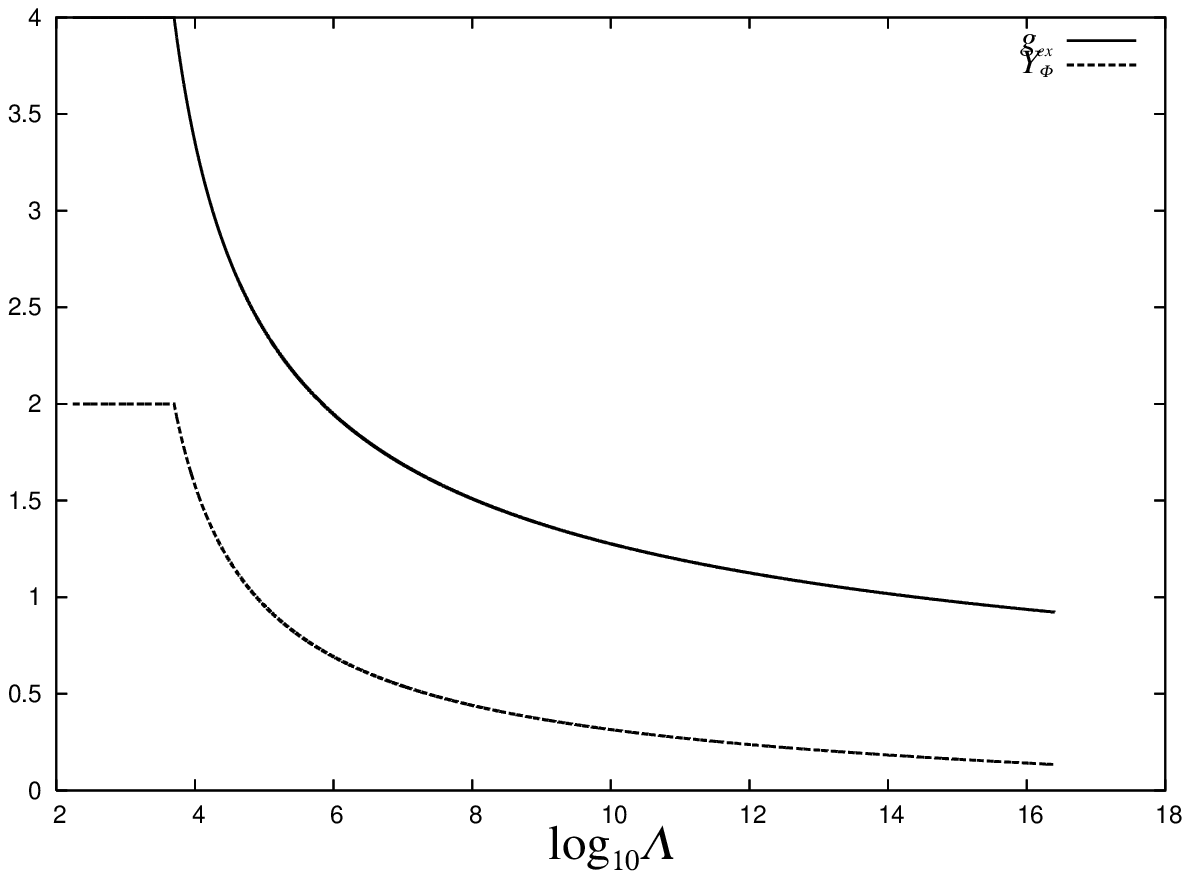}
\caption{\label{fig6} 
The RG flows of the gauge couplings $g_{ex}$
and the Yukawa coupling $y_{\Phi}$ in the extra sector
are shown in the case of 
$\alpha_{ex} \sim \alpha^*_{ex}$ and 
$\alpha_{y_{\Phi}} = 1/\pi$ 
$(y_{\Phi}=2)$ at the
decoupling scale 5 TeV.
}
\includegraphics[width=0.45\textwidth]{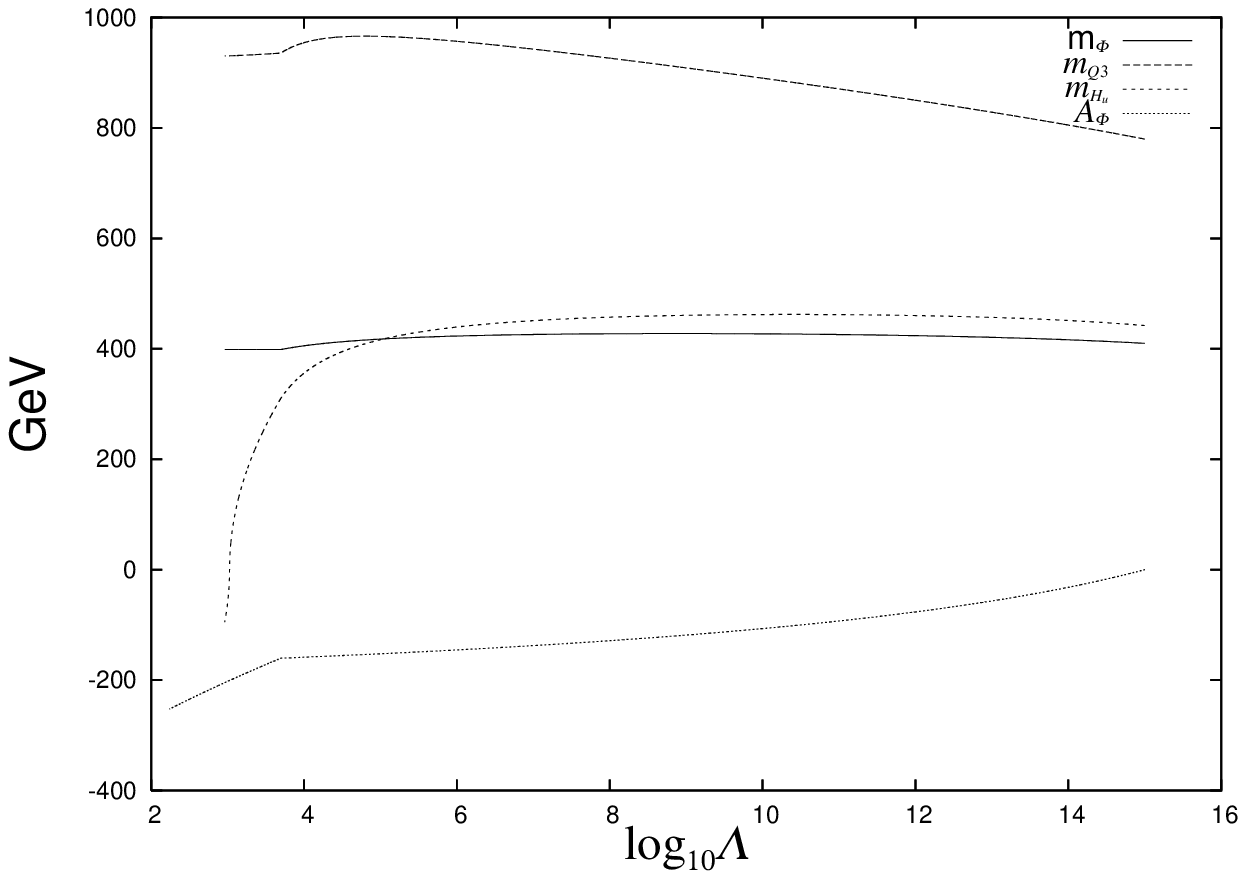}
\caption{\label{fig7} 
The RG flows of the soft scalar masses
$m_{\Phi}$, $m_{\phi}, m^2_{H_u}/|m_{H_u}|$
realized in the Case (2) are shown.
The figure shows also the RG flow of $A_{\Phi}$
which is connected with the flow of $A_{t}$ 
of the MSSM at the decoupling scale.
}

\end{figure}

Now it is almost obvious that the fine-tuning
problem is removed in this model, since
the $\mu$-parameter required in order to 
realize the EWSB is also the weak scale.
Fig.~8 and Fig.~9 present the minimum degree of fine-tuning
$\Delta_{\mu^2}$ required in the extra $SU(3)$ model for 
$\tan \beta =10$  and $5$ respectively.
(The data points with enough small degrees are not shown,
since the model is safe from the fine-tuning anyway.)
In the both figures, the upper lines show the minimum
degrees to satisfy the lightest Higgs mass bound and
the lower lines show the minimum
degrees to satisfy the selectron mass bound.

These results tell us the followings.
The lightest Higgs mass bound
always imposes the dominant constraint.
However, the extra $SU(3)$ models with a large value of
$\tan \beta$ is almost free from the severe
fine-tuning for any messenger scale. 
It is also shown that $\mu$-parameter becomes smaller
as the messenger scale goes higher, and is 
comparable with the weak scale for the high messenger
scale GMSB.

Even in the  case of $\tan \beta=5$, 
the degree of fine-tuning is reduced significantly 
compared with the ordinary MSSM.
However, it is found that
the minimum degree of fine-tuning increases rapidly
for $\tan \beta < 5$. 
This is because the required stop mass is so heavy 
({\it e.g.} $2.5$TeV for $\tan \beta=3$) 
as to induce a huge correction to the Higgs soft mass
below the decoupling scale.
The cancellation of the SM correction becomes also
significant with such a large supersymmetry breaking 
scale anyway.
Thus we may conclude that the natural models are 
obtained for a $\tan \beta$ more than 5 or so.

\begin{figure}[htb]
\includegraphics[width=0.45\textwidth]{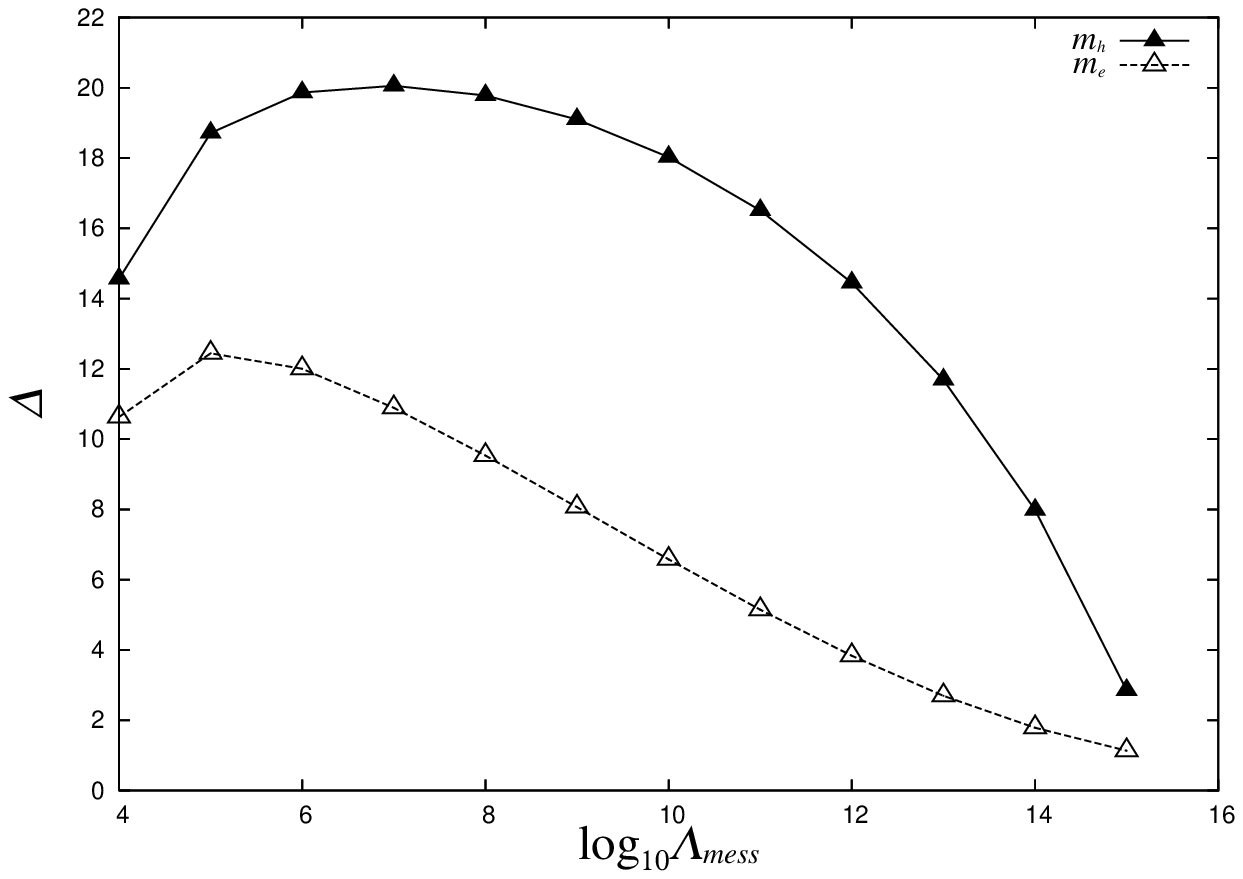}
\caption{\label{fig8} 
The minimum degrees of fine-tuning required in the
extra $SU(3)$ models with various messenger scales
are shown in the case of $\tan \beta = 10$.
The conditions for the lines are the same as 
in Fig.~1.
The data points for $\Lambda_{mess}=10^{16}$GeV are
not given, since the EWSB solution is not realized.
}
\includegraphics[width=0.45\textwidth]{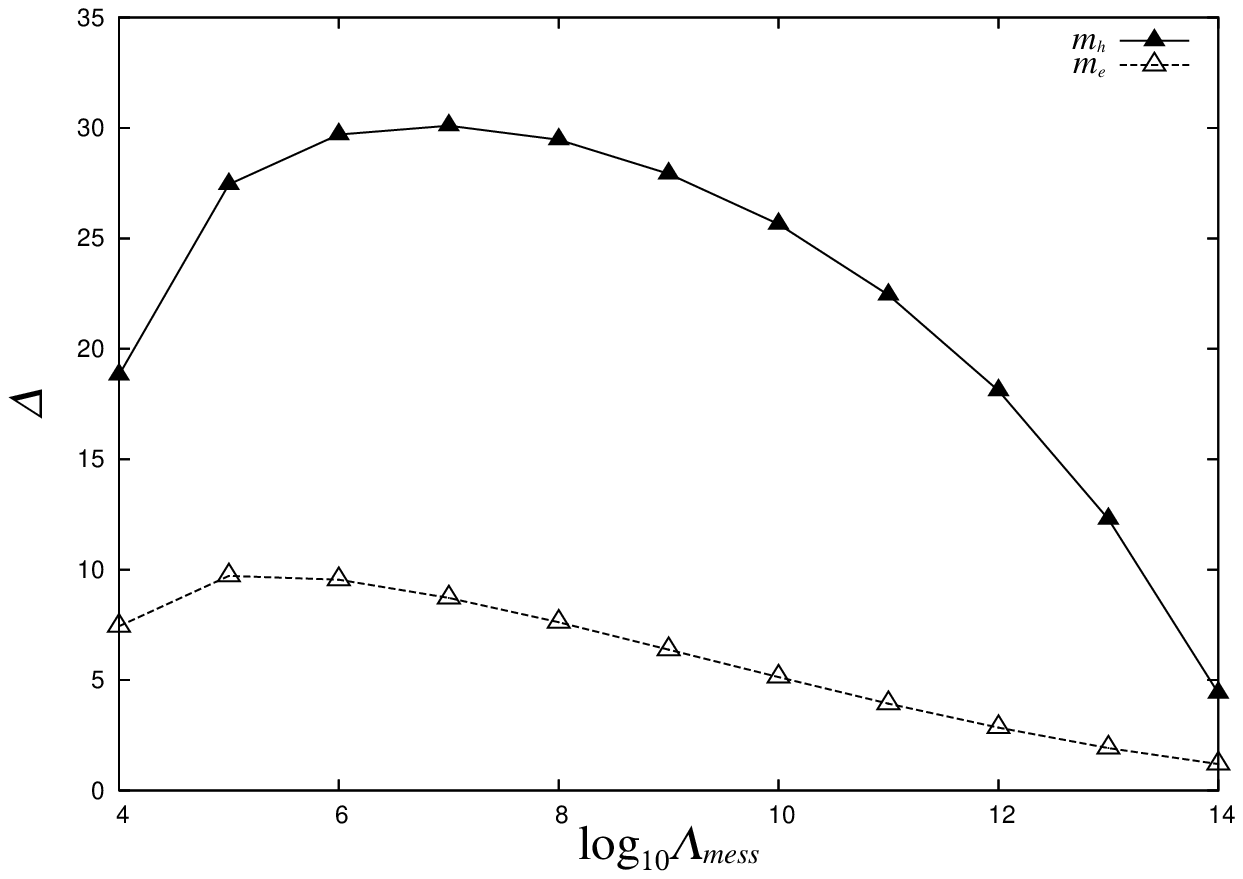}
\caption{\label{fig9} 
The minimum degrees of fine-tuning
in the case of $\tan \beta = 5$ are shown in the
same way as Fig.~8.
}
\end{figure}

\section{Extension to the next-to-minimal model}

\subsection{Fine-tuning problem in the NMSSM}

The $\mu$-parameter as well as the B-parameter
of the Higgs sector must be the same order of 
the soft supersymmetry 
breaking parameters in order to realize the EWSB.
Therefore it would be natural to think that the 
$\mu$-parameter also has some origin related with the 
supersymmetry breaking.
In the GMSB models, this problem appears to be a harder
puzzle, since the generated B-parameter is 
given naively by 
$B_S = \langle F_S \rangle/\langle S \rangle$,
which is too large for the Higgs potential. 
The simplest mechanism to solve this problem
would be the NMSSM \cite{NMSSM},
while there have been proposed some other mechanisms
\cite{muproblem}.

The NMSSM introduces  a SM-gauge singlet field $N$
replacing the $\mu$ term with  the Yukawa
term in the superpotential given by (\ref{NMSSM}).
Then the $\mu$-parameter may be generated through 
a non-vanishing vev of $N$ as 
$\mu = \lambda_H \langle N \rangle$.
However this vev breaks the global $U(1)_{PQ}$ symmetry
of the superpotential spontaneously and, therefore,
gives rise to the EW-scale axion, which is
excluded experimentally \cite{HW}.
Here we avoid this problem by breaking the  $U(1)_{PQ}$ 
symmetry explicitly with introducing an additional 
term,
\be
\delta W = \frac{k}{3} N^3,
\ee
where the parameter $k$ is supposed to be
comparable with $\lambda_H$.

We should mention here the new possibility for 
solving the fine-tuning problem proposed 
recently in Ref.~\cite{DG}.
When the pseudo-NG boson $a$ is very light, decay of
the SM-like light Higgs boson $h$ to $a$, 
$h \rightarrow aa$, can be
dominant over decay to the bottom quarks, 
$h \rightarrow b\bar{b}$.
Then the Higgs may be missed and the mass can be 
lower than the bound of $114$GeV.
However this possibility opens only when
the coupling $k$ is rather small to make
the pseudo-NG boson of the broken 
$U(1)_{PQ}$ sufficiently light.
Then a large A-parameter is required to raise the
singlino and higgsinos above LEP bounds \cite{EWPT},
which is not realized in the GMSB.
Therefore we do not consider the cases with
a small $k$ in our discussion.

So far the fine-tuning required in the NMSSM has been
investigated minutely in Refs.~\cite{AG,dFM}.
Here let us review the essence of their argument
and give an analytic expression for the fine-tuning
degrees by using somewhat rough approximations.

The parameters which we can tune freely are the Yukawa
couplings $\lambda_H$ and $k$ only in the GMSB.
Note that $\mu$ or $\tan \beta=v_u/v_d$ is not a 
free parameter any more, but is determined dynamically
by the  scalar potential.
In order to discuss the fine-tuning, we must express
the solution for $v^2$ in terms of the Yukawa couplings
first. Then we examine stability of the solution
under a small variation of the couplings.

\begin{widetext}
The potential of  the NMSSM model is written down as
\bea
V &=&
\left| \lambda_H H_d H_u - k N^2 \right|^2 
+ \lambda_H^2 |N|^2
\left( |H_d|^2 + |H_u|^2 \right)
+ ~\frac{g^2}{8} \left(
H_d^{\dagger} \tau^a H_d + H_u^{\dagger} \tau^a H_u
\right)^2
+ ~\frac{{g'}^2}{8} \left(
|H_d|^2 - |H_u|^2 \right)^2  \nn \\
& & 
+ ~m^2_{H_u} |H_u|^2 + m^2_{H_d} |H_d|^2 + m^2_{N} |N|^2 
- ~(\lambda_H A_{\lambda_H} N H_u H_d + \mbox{h.c.} )
- \left(\frac{k}{3} A_k N^3 +  \mbox{h.c.} \right) + \cdots,
\eea
where the ellipsis denotes higher order terms.
Then the minimization conditions for the neutral
scalar components are found to be
\bea
& &
\lambda_H^2 x^2 =
- \frac{g^2 + g'^2}{8}v^2 +
\frac{m^2_{H_d} - m^2_{H_u} \tan^2 \beta}{\tan^2 \beta -1} 
\label{NMSSMeq1} \\
& &
A_{\lambda_H} \lambda_H x =
\left( 
m^2_{H_d} + m^2_{H_u} + 2 \lambda_H^2 x^2
\right)\frac{\sin 2\beta}{2} 
\label{NMSSMeq2} \\
& & 
2 k^2 x^2 =
\lambda_H v^2 (k \sin 2\beta - \lambda_H) - m^2_N
+ A_{\lambda_H} \lambda_H v^2 \frac{\sin 2 \beta}{2 x}
+ k A_k x,
\label{NMSSMeq3}
\eea
where  $v_u = \langle H_u \rangle$, 
$v_d = \langle H_d \rangle$, 
$(v^2 = v_u^2 + v_d^2)$ and 
$x = \langle N \rangle$.
It is seen that the first and the second equations are 
identical with the conditions in the MSSM by 
replacing as $\lambda_H x = \mu$ and $A_{\lambda_H} = B$.

Here we shall approximate the solutions of the above
equations by performing some simplifications.
{}First we neglect the A-parameters, 
since $A_{\lambda_H}$ is much smaller than $v$ and $A_k$ is
negligible in the GMSB irrespectively of the messenger
scale.
Next we simplify the conditions further by setting
as $m^2_{H_u} + m^2_{H_d} \sim m^2_{H_u}$ and
$\sin 2 \beta \sim 2/\tan \beta$.
Although these simplifications are not always good 
quantitatively, we may find roughly how strong 
fine-tuning is required.
Then the equations are reduced to be
\bea
& &
\mu^2 + \frac{c}{2} v^2 + m^2_{H_u} = 0, 
\label{cond1}\\
& &
\kappa \tan \beta \mu^2 + (c - \lambda_H^2) v^2 + m^2_{H_u}
=0, 
\label{cond2}\\
& &
2 \kappa^2 \mu^2 + \lambda_H^2 v^2 
\left(
1 - \frac{2\kappa}{\tan \beta}
\right) + m^2_N = 0,
\label{cond3}
\eea
where $\mu^2 = \lambda_H^2 x^2$, 
$c = (g^2+{g'}^2)/4$ and $\kappa=k/\lambda_H$.
After eliminating $\mu^2$ from these equations,
the conditions are reduced further to 
\bea
F_1
&\equiv&
- \left( 1 - \kappa \tan \beta \right)m^2_{H_u}
- \left( 1 - \frac{\kappa}{2} \tan \beta \right) \alpha v^2
+ \lambda_H^2 v^2 = 0 
\label{F1}\\
F_2
&\equiv&
2 \kappa^2 \left( -m^2_{H_u} - \frac{\alpha}{2} v^2 \right)
+ \lambda_H^2 v^2 
\left( 1 - \frac{2 \kappa}{\tan \beta} \right)
+ m^2_N = 0.
\label{F2}
\eea

As seen before, the soft scalar mass $|m_{H_u}|$
is far larger than the weak scale in the GMSB.
Therefore we first consider the conditions so that 
the Yukawa couplings $\lambda_H$ and $k$ should satisfy
in order to have $v^2 \ll |m^2_{H_u}|$ as a
solution.
However, it has been known that the 
phenomenologically viable spectrum cannot be 
obtained in the GMSB \cite{AG,dFM}.
The reason is that the Yukawa interaction with 
$\lambda_H$ itself cannot drive the soft mass for 
$N$, $m^2_N$, to be sufficiently negative
at the weak scale. 

{}From the equation (\ref{cond1}), it is found
that $\mu^2 \sim |m^2_{H_u}| \gg v^2$.
Meanwhile the equation (\ref{cond3}) may be
rewritten as \cite{AG}
\bea
-2 \kappa^2 \mu^2 & = &  
\lambda^2_H v^2
\left(
1 - \frac{2\kappa}{\tan \beta}
\right) + m^2_N  \nn \\
&\simeq& \lambda_H^2
\left[
v^2 \left(
1 - \frac{2\kappa}{\tan \beta}
\right)
 - \frac{1}{4\pi^2}
\left(
m^2_{H_u}(\Lambda_{mess}) + m^2_{H_d}(\Lambda_{mess})
\right)
\log \frac{\Lambda_{mess}}{m_{H_d}}
\right],
\eea
where we evaluated $m_N^2$ by using the one-loop
perturbation.
However this equation cannot be satisfied even with
a very large messenger scale $\Lambda_{mess}$,
since $\kappa$ cannot be taken so small.
Therefore the NMSSM is often modified to have also 
a pair of extra vectorlike color-triplet fields
coupled with the singlet $N$ \cite{GMSB,muproblem}.
Then the low energy scalar potential can lead 
to a viable EWSB with $m^2_N$ sufficiently negative.
In Refs. \cite{AG,dFM}, the fine-tuning problem 
has been investigated in the NMSSM model with the 
extra ``quarks'' and was found to be more serious
than in the MSSM.
\end{widetext}

Hence we may consider the cases with
$|m^2_N| \sim |m^2_{H_u}| \gg v^2$.
Then it is seen from Eqs.~(\ref{F1}) and (\ref{F2})
that such a hierarchical difference among
the mass scales can appear, 
only when  the ratio of the couplings 
$\lambda_H$ and $k$ happen to be
\be
\kappa^2 = \frac{k^2}{\lambda_H^2} 
\sim \frac{|m^2_N|}{2|m^2_{H_u}|}.
\ee
If the parameter $\kappa$ deviates from this relation,
then the solution of $v^2$ becomes comparable with 
$|m^2_{H_u}|$ immediately. This is the origin of
the strong fine-tuning.
We also note that 
$\tan \beta$ of the solution is also fixed to be
\be
\tan \beta \sim 1/\kappa
\ee
approximately. This is the reason why $\tan \beta$ cannot
be large in the NMSSM models in general.

The response of the solutions ($v^2, \tan \beta$) 
under a slight shift of {\it e.g.} $\kappa$ 
satisfies the following relations,
\bea
& &
\kappa \frac{\partial F_1}{\partial \kappa} 
+ \kappa \frac{\partial v^2}{\partial \kappa}
\frac{\partial F_1}{\partial v^2}  
+ \kappa \frac{\partial \tan \beta}{\partial \kappa}
\frac{\partial F_1}{\partial \tan \beta} =0,  \\
& &
\kappa \frac{\partial F_2}{\partial \kappa} 
+ \kappa \frac{\partial v^2}{\partial \kappa}
\frac{\partial F_2}{\partial v^2}  
+ \kappa \frac{\partial \tan \beta}{\partial \kappa}
\frac{\partial F_2}{\partial \tan \beta} =0.
\eea  
By solving these equations, we may evaluate 
the degree of fine-tuning in order to realize the
specific solution of $v^2$.
Since we consider the solutions with
$v^2 \ll |m^2_N| \sim |m^2_{H_u}|$,
the degree is found to be approximated as
\bea
\Delta_{\kappa}(v^2) 
&=& 
\left|
\frac{\kappa}{v^2}
\frac{ \partial v^2}{\partial \kappa}
\right| \nn \\
&\sim& 
\left|
\frac{2 \kappa^2}{\kappa^2 - 
\lambda_H^2 (1 - 2\kappa^2)/c
}
\right|
\frac{2 |m^2_{H_u}|}{ M^2_Z}.
\label{estimation}
\eea
Although we should examine also the deviation 
of $\lambda_H$, the solutions are found to be 
rather stable.
Here we note that the quantity $2|m^2_{H_u}|/M^2_Z$
is nothing
but the degree of fine-tuning for the MSSM case.
This is the analytic expression for the
fine-tuning index for the NMSSM in the GMSB.
We may certify the validity of this formula
by solving the coupled equations numerically,
(\ref{NMSSMeq1}), (\ref{NMSSMeq2}), (\ref{NMSSMeq3}),
with incorporating all of the parameters given by the
GMSB framework.
It is found to be good as long as 
$|m^2_{H_u}|, |m^2_N| \gg v^2$.

The Yukawa coupling $\lambda_H$ cannot be taken more than
$0.5$, otherwise the running coupling hits the
Landau pole well before the GUT scale.
Therefore, unless the denominator of the index
given by (\ref{estimation})
happen to be small, the fine-tuning
in the NMSSM is the same as that obtained in the 
MSSM with the same $\tan \beta$
up to the factor of O(1) \cite{dFM}.
Consequently, the fine-tuning required in the NMSSM models
is found to be no better than in the MSSM.

The crucial difference between the MSSM and the NMSSM
is the value of $\tan \beta$.
However, the term of $\lambda_H N H_u H_d$ in the
superpotential lifts up the lightest Higgs mass bound
for the small $\tan \beta$ case as
\be
m_{h^0}^2 \geq
\left(
M_Z^2 \cos^2 2 \beta + \lambda^2_H v^2 \sin^2 2 \beta
\right)
+ (\mbox{loop}).
\label{NMSSMmassbound}
\ee
Therefore the constraint on the stop masses from the 
Higgs mass bound becomes almost independent of $\tan \beta$
for $\lambda_H \sim 0.5$.
Unless the coupling $\lambda_H$ is given to be so large, 
the fine-tuning is very severe in the NMSSM models
\cite{fathiggs}.

\subsection{Extension of the extra $SU(3)$ model}

Lastly we consider the fine-tuning in 
the extra $SU(3)$ model modified 
so that the supersymmetric mass parameters are generated
through vev of a singlet $N$.
The superpotential is given explicitly by
\bea
W &=& y_t Q \bar{u} H_u 
+ \lambda_H N H_d H_u - \frac{1}{3} k N^3 \nn \\
& & 
+ y_{\Phi} \Phi \bar{\phi} H_u 
+ \lambda_{\Phi} N \Phi \bar{\Phi} \nn \\
& &
+ \lambda_{\phi} N \phi \bar{\phi}
+ \lambda_{\Omega} N \Omega \bar{\Omega}.
\label{WofNMSSM}
\eea
The decoupling masses of
the extra vectorlike fields as well as 
the $\mu$ parameter are induced through the singlet
vev as $M_{\Phi} = \lambda_{\phi} \langle N \rangle$,
{\it etc.}.
Here, however, these masses should be much larger
than $\mu$, {\it e.g.} a few TeV.
It is interesting that such a hierarchy may be also
generated by the strong gauge dynamics in the extra
sector.
The Yukawa couplings, $\lambda_{\Phi}$, 
$\lambda_{\Phi}$ and $\lambda_{\Omega}$, are also
enhanced by the strong gauge coupling $\alpha_{ex}$
similarly to $y_{\Phi}$, while $\lambda_H$ is not
enhanced.
Therefore a large discrepancy among these couplings
arises at the low energy scale naturally.

It may be an advantage for the extra $SU(3)$ model 
to contain the extra vectorlike fields coupled 
with $N$ from the beginning.
Owing to these couplings, 
the soft scalar mass $m^2_N$ can be driven to be
sufficiently negative at the low energy scale.
In practice, the viable models for the EWSB 
are found to be obtained for appropriate choices of
the Yukawa couplings.

The main difference between the present model 
and the conventional NMSSM model examined 
in Refs. \cite{AG,dFM} is that the size of the
soft scalar mass $|m^2_{H_u}|$ obtained
at the weak scale is not so enhanced by the
large stop masses. 
That is because the extended model also 
enjoys the nature 
that the radiative  correction to $m^2_{H_u}$
can be effectively cutoff around the decoupling
scale.
On the other hand, however, 
the problem is that the coupling $\lambda_H$ cannot
be taken so large as $0.5$.
First the Yukawa couplings $\lambda_{\Phi}$ and 
$\lambda_{\phi}$ should be less than $0.5$,
otherwise these Yukawa interactions induce
a sizable soft mass $|m_N|$, and, therefore,
the solution of $\tan \beta$ becomes too small.
Then the coupling $\lambda_H$
should be smaller than about $0.05$ to make 
the mass ratio 
$\lambda_{\Phi}/\lambda_H = M_{\Phi}/\mu$
sufficiently large,
otherwise the decoupling scale of the extra
sector is too low to pass the EW precision tests.

Thus, unfortunately the above extension of the
extra $SU(3)$ model to the NMSSM type appears
to be unsuccessful.
The fine-tuning is so severe as in the MSSM with
a small $\tan \beta$ as seen 
in the previous section.
However, this situation can be improved remarkably, 
if the Yukawa coupling $\lambda_H$ is allowed to 
be somewhat large as $0.3$-$0.5$.
Then some other mechanisms to generate
the decoupling scale of a few TeV is required.
We leave this problem in the NMSSM model to the 
future studies.

\section{Conclusion}

In this paper we first studied the degree of
fine-tuning in the GMSB with an arbitrary messenger
scale. Specially we took  the lightest
Higgs mass bound by LEP into consideration
by using the 2 loop approximation.
The spectra of all soft supersymmetry breaking
parameters obtained at the weak scale
are completely fixed up to their overall scale $B_S$.
Therefore we examined the fine-tuning with respect to the 
$\mu$-parameter required so as to 
realize the EWSB.

Then it was found that the constraint on the 
soft masses by the Higgs mass bound is far more
restrictive than by the selectron mass bound 
for entire range of the messenger scale, 
and the fine-tuning stronger than 
a few \% is required at the very best.
The degree becomes bigger as the messenger scale is
lowered, because the A-parameter, $A_t$, is given
to be small.
Moreover, the model with a small value of $\tan \beta$ 
($\leq 5$) needs an extreme fine-tuning of the
$\mu$-parameter.

In order to solve this problem, we studied a new 
scenario in which the top Yukawa 
coupling is induced at TeV scale through mass mixing
with some extra matter fields coupled with the Higgs
field. 
We also applied the partial GMSB mechanism 
that the extra gauge sector is sequestered from 
the supersymmetry breaking sector.
In the explicit model of the Case (2) 
presented in section IV,
the radiative correction
to the Higgs soft mass can be cutoff effectively
at the TeV scale.
Then the Higgs soft mass is found to be comparable 
with the weak scale and, therefore, renders
the fine-tuning of the $\mu$-parameter unnecessary,
even if the stop masses are quite large.
In practice, we found that the natural models are 
obtained for $\tan \beta$ more than 5 or so.

We also examined the oblique corrections induced by the
mixing with the extra fields and checked that the
models are safe as long as the extra particles are
heavier than 1TeV. 
The feature of this model is that the Higgs scalars 
and the Higgsinos are so light as the weak scale, while
the squarks are quite heavy.
Moreover, the sleptons can be fairly heavier 
than the bino for a high messenger scale.
This is because the running behaviors of the SM gauge
couplings are modified by the presence of the 
extra fields.

The extension to the NMSSM is supposed to be the easiest
way to avoid the $\mu$-problem in the GMSB framework.
It has been known, however, that the additional correction
by the extra ``quarks'' is necessary to make a viable model.
In the extra $SU(3)$ model, the fields in the extra gauge 
sector coupled with the singlet field play this role and
the realistic EWSB can take place.

Unfortunately, the extended model turns out to suffer 
fine-tuning, because $\tan \beta$ determined by the EWSB
vacuum is small. 
In the explicit model the Yukawa coupling of the singlet
with the Higgs field $\lambda_H$ is taken to be rather
small in order to make the decoupling mass heavy. 
However, if  $\lambda_H$ can be enhanced,
the fine-tuning problem may be resolved just as in the
MSSM with large $\tan \beta$.
The improvement of the NMSSM model in this way is left
for the future study.

\section*{Acknowledgements}

T.~K.\ and H.~T. are supported in part by the Grants-in-Aid for 
Scientific Research No.~17540251 and No.~40192653 respectively
from the Ministry of Education, Science, Sports and 
Culture, Japan.
T.~K.\ is also supported in part by
the Grant-in-Aid for the 21st Century COE
``The Center for Diversity and Universality in Physics".

\section*{Note Added}

As this paper was prepared, we found that
a comprehensive paper
by R.~Kitano and Y.~Nomura \cite{KN}
appeared and our study given in section II 
overlaps in part.



\begin{thebibliography}{99}
\bibitem{BG}
R.~Barbieri and G.~F.~Giudice, 
Nucl.~Phys.~B {\bf 306}, 63 (1988).

\bibitem{finetune}
P.H.~Chankowski, J.R.~Ellis and S.~Pokorski,
Phys.~Lett.~B {\bf 423}, 327 (1998);
Nucl.~Phys.~B {\bf 544}, 39 (1999);
G.L.~Kane and S.F.~King, 
Phys.~Lett.~B {\bf 451}, 113 (1999);
M.~Bastero-Gil, G.L.~Kane and S.F.~King,
Phys.~Lett.~B {\bf 474}, 103 (2000).

\bibitem{LEPII}
[ALEPH, DELPHI, L3 and OPAL Collaborations],
Phys.~Lett.~B {\bf 551}, 146 (2003).

\bibitem{massbound}
Y.~Okada, M.~Yamaguchi and T.~Yanagida,
Phys.\ Lett.\ B {\bf 262}, 54 (1991);
H.E.~Haber and R.~Hempfling,
Phys.\ Rev.\ Lett.\  {\bf 66}, 1815 (1991);
J.R.~Ellis, G.~Ridolfi and F.~Zwirner,
Phys.\ Lett.\ B {\bf 262}, 477 (1991).

\bibitem{2loop}
M.~Carena, M.~Quir\'{o}s and C.E.M.~Wagner,
Nucl.\ Phys.\ B {\bf 461}, 407 (1996);
M.~Carena, H.E.~Haber, S.~Heinemeyer,
W.~Hollik, C.E.M.~Wagner and G.~Weiglein
Nucl.\ Phys.\ B {\bf 580}, 29 (2000).

\bibitem{radSB}
K.~Inoue, A.~Kakuto, H.~Komatsu and S.~Takeshita,
Prog.\ Theor.\ Phys.\  {\bf 67}, 1889 (1982);
Prog.\ Theor.\ Phys.\  {\bf 68}, 927 (1982)
[Erratum-ibid.\  {\bf 70}, 330 (1983)];
Prog.\ Theor.\ Phys.\  {\bf 71}, 413 (1984);
L.E.~Ibanez and G.G.~Ross,
Phys.\ Lett.\ B {\bf 110}, 215 (1982);
L.~Alvarez-Gaume, M.~Claudson and M.B.~Wise,
Nucl.\ Phys.\ B {\bf 207}, 96 (1982).

\bibitem{darkmatter}
See for a review, {\it e.g.},
J.L.~Feng, 
arXiv:hep-ph/0405215;
see also 
N.~Arkani-Hamed, A.~Delgado and G.F.~Giudice,
arXiv:hep-ph/0601041
and references therein.

\bibitem{GMSB}
See for a review, {\it e.g.},
G.F.~Giudice and R.~Rattazzi,
Phys.\ Rept.\ {\bf 322} 419 (1999)
and references therein.

\bibitem{AG}
K.~Agashe and M.~Graesser,
Nucl.~Phys.~B {\bf 507}, 3 (1997). 

\bibitem{dFM}
A.~de~Gouv\^ea, A.~Friedland and H.~Murayama,
Phys,~Rev. {\bf D 57}, 5676 (1998). 

\bibitem{muproblem}
M.~Dine and A.~Nelson, 
Phys,~Rev. {\bf D 48}, 1277 (1993);
M.~Dine, A.~Nelson and Y.~Shirman, 
Phys,~Rev. {\bf D 51}, 1362 (1995);
M.~Dine, A.~Nelson, Y.~Nir and Y.~Shirman, 
Phys,~Rev. {\bf D 53}, 2658 (1996);
G.~Dvali, G.F.~Giudice and A.~Pomarol,
Nucl.~Phys.~B {\bf 478}, 31 (1996). 

\bibitem{NMSSM}
P.~Fayet, 
Nucl.~Phys.~B {\bf 90}, 104 (1975);
H.P.~Nilles, M.~Srednicki and D.~Wyler,
Phys.\ Lett.\ B {\bf 120}, 346 (1983);
J.M.~Frere, D.R.T.~Jones and S.~Raby,
Nucl.~Phys.~B {\bf 222}, 11 (1983);
J.P.~Derendinger and C.A.~Savoy,
Nucl.~Phys.~B {\bf 237}, 307 (1984);
J.R.~Ellis, J.F.~Gunion, H.E.~Haber,
L.~Roszkowski and F.~Zwirner,
Phys,~Rev. {\bf D 39}, 844 (1989).

\bibitem{Casas}
A.~Brignole, J.A.~Casas, J.R.~Espinosa and
I.~Navarro,
Nucl.~Phys.~B {\bf 666}, 105 (2003);
J.A.~Casas, J.R.~Espinosa and I.~Hidalgo,
JHEP {\bf 0401} 008  (2004).

\bibitem{fathiggs}
R.~Harnik, G.D.~Kribs, D.T.~Larson and H.~Murayama,
Phys.~Rev. {\bf D70},  015002 (2004);
S.~Chang, C.~Kalic and R.~Muhbubani,
Phys.~Rev. {\bf D71},  015003 (2005);
A.~Birkedal, Z.~Chacko and Y.~Nomura,
Phys.~Rev. {\bf D71},  015006 (2005);
A.~Delgado and T.M.P.~Tait,
JHEP {\bf 0507},  023 (2005). 


\bibitem{Delgado}
P.~Batra, A.~Delgado, D.E.~Kaplan and T.M.~P.~Tait,
JHEP {\bf 0402}, 043 (2004); 
JHEP {\bf 0406},  032 (2004).


\bibitem{supersoft}
P.J.~Fox, A.E.~Nelson and N.~Weiner,
JHEP 0208 (2002) 035;
Z.~Chacko, P.J.~Fox and H.~Murayama,
Nucl.~Phys.~B {\bf 706}, 53 (2005).


\bibitem{superLH}
A.~Birkedal, Z.~Chacko and  M.K.~Gaillard, 
JHEP {\bf 0410}, 036 (2004);
P.H.~Chankowski, A.~Falkowski, S.~Pokorski
and J.~Wagner,
Phys.\ Lett.\ B {\bf 598}, 252 (2004);
Z.~Berezhiani, P.H.~Chankowski, A.~Falkowski
and S.~Pokorski,
arXiv:hep-ph/0509311;
T.~Roy and M.~Schmaltz,
JHEP {\bf 0601}, 149 (2006);
C.~Cs\'{a}ki, G.~Marandella, Y.~Shirman and 
A.~Strumia,
Phys.~Rev.~D {\bf 73}, 035006 (2006).

\bibitem{KT}
T.~Kobayashi and H.~Terao,
JHEP {\bf 0407}, 026 (2004).

\bibitem{KNT}
T.~Kobayashi, H.~Nakano and H.~Terao,
Phys.~Rev.~D {\bf 71}, 115009 (2005).


\bibitem{NPT}
Y.~Nomura, D.~Poland and B.~Tweedie,
Nucl.~Phys.~ B {\bf 745}, 29 (2006).

\bibitem{KKLT}
K.~Choi, K.S.~Jeong, T.~Kobayashi and K.i.~Okumura,
Phys.\ Lett.\ B {\bf 633}, 355 (2006);
R.~Kitano and Y.~Nomura,
Phys.\ Lett.\ B {\bf 631}, 58 (2006);

\bibitem{DK}
R.~Dermisek and H.D.~Kim,
arXiv:hep-ph/0601036.

\bibitem{KN}
R.~Kitano and Y.~Nomura,
arXiv:hep-ph/0602096.

\bibitem{twinhiggs}
A.~Falkowskoi, S.~Pokorski and M.~Schmaltz,
arXiv:hep-ph/0604066;
S.~Chang, L.~Hall and N.~Weiner,
arXiv:hep-ph/0604076.

\bibitem{focus}
J.L.~Feng, K.T.~Matchev and T.~Moroi,
Phys.~Rev.~Lett. {\bf 84}, 2322 (2000);
Phys.~Rev.~D {\bf 61}, 075005 (2000);
J.L.~Feng, K.T.~Matchev and F.~Wilczek,
Phys.~Lett.~B {\bf 482}, 388 (2000).

\bibitem{GR}
G.F.~Giudice and R.~Rattazzi,
Nucl.~Phys.~B {\bf 511}, 25 (1998).

\bibitem{seiberg}
N.~Seiberg, 
Nucl.~Phys.~B {\bf 435}, 129 (1995);
K.~Intrilligator and N.~Seiberg, 
Nucl.~Phys.~Proc.~Suppl. {\bf 45BC}, 1 (1996).

\bibitem{EWPT}
The LEP Collaborations ALEPH, DELPHI, L3,
OPAL and the LEP Electroweak Working
Group, 
arXiv:hep-ph/0511027.


\bibitem{topseesaw}
B.A.~Dobrescu and C.T.~Hill,
Phys.~Rev.~Lett. {\bf 81}, 2634 (1998);
R.S.~Chivukula, B.A.~Dobrescu, H.~Georgi and C.T.~Hill,
Phys.~Rev.~D {\bf 59}, 075003 (1999);
H.~Collins, A.K.~Grant and H.~Georgi,
Phys.~Rev.~D {\bf 61},  055002 (2000).

\bibitem{LH}
See for a review, {\it e.g.},
M.~Schmaltz and D.~Tucker-Smith,
arXiv:hep-ph/0502182.

\bibitem{sumrule}
A.E.~Nelson and M.J.~Strassler,
JHEP {\bf 0009}, 030 (2000);
JHEP {\bf 0207}, 021 (2002).
T.~Kobayashi and H.~Terao,
Phys.~Rev.~D {\bf 64}, 075003 (2001);
T.~Kobayashi, H.~Nakano and H.~Terao, 
Phys.~Rev.~D {\bf 65}, 015006 (2002);
T.~Kobayashi, H.~Nakano, T.~Noguchi and H.~Terao, 
Phys.~Rev.~D {\bf 66}, 095011 (2002).

\bibitem{DG}
R.~Dermisek and J.~F.~Gunion,
Phys.~Rev.~Lett. {\bf 95} 041801 (2005);
S.~Chang, P.~J.~Fox and N.~Weiner,
arXiv:hep-ph/0511250;
P.~C.~Schuster and N.~Toro,
arXiv:hep-ph/0512189.

\bibitem{HW}
See for a recent analysis, {\it e.g.},
L.J.~Hall and T.~Watari,
Phys.~Rev.~ {\bf D70}, 115001 (2004).


\end{thebibliography}
\end{document}